%% file: Paper.tex
\documentclass{aa}

\usepackage{txfonts}
\usepackage{graphicx}
\usepackage{natbib}
\bibpunct{(}{)}{;}{a}{}{,}            
\usepackage[figuresright]{rotating}
\usepackage[english]{babel}
\usepackage{aalongtable}
\graphicspath{{figures/}}

\begin{document}

\title{Variability and periodicity of field M dwarfs revealed by multichannel monitoring}
\author{Boris Rockenfeller \and
  Coryn~A.~L. Bailer-Jones\thanks{\emph{Send offprint requests
  to}: calj@mpia.de} \and Reinhard Mundt}
\institute{Max-Planck-Institut f\"ur Astronomie, K\"onigstuhl 17, 69117
  Heidelberg, Germany}
\date{Received 6 September 2005 / Accepted 15 November 2005}

\authorrunning{Rockenfeller et al.}
\titlerunning{Variability and periodicity of field M dwarfs}

\abstract{We present simultaneous, multiband photometric monitoring
of $19$ field dwarfs covering most of the M spectral sequence
(M2--M9). Significant variability was found in seven objects in at
least one out of the three channels I, R and G. Periodic
variability was tested with a CLEAN power spectral analysis. Two
objects, LHS370 (M5V) and 2M1707+64 (M9V), show periods of $5.9\pm
2.0$ and $3.65\pm 0.1$ hours respectively. On account of the agreement
with the typical values of $v\sin i$ published for M dwarfs
\citep{Mohanty2003}, we claim these to be the objects' rotation
periods. Three further objects show possible periods of a few
hours. Comparing the variability amplitude in each channel with
predictions based on the synthetic spectra of \citet{Allard2001}, we
investigated the source of variability in LHS370 and 2M1707+64. For
the latter, we find evidence for the presence of magnetically-induced
cool spots at a temperature contrast of $4-8\,\%$, with a projected
surface coverage factor of less than 0.075.
Moreover, we can rule out dust clouds (as represented by the COND or DUSTY
models) as the cause of the variability. No conclusion can be drawn in
the case of LHS370. Comparing the frequency of occurrence of
variability in this and various L dwarf samples published
over the past few years, we find that variability is more common in
field L~dwarfs than in field M~dwarfs (for amplitudes larger than $0.005\,{\rm
mag}$ on timescales of $0.5$ to $20$ hours). Using the homogeneous
data sets of this work and \citet{Bailer-Jones2001}, we find fractions
of variable objects of $0.21\pm 0.11$ among field M dwarfs and
$0.70\pm 0.26$ among field L dwarfs (and $0.29\pm 0.13$, $0.48\pm
0.12$ respectively if we take into account a larger yet more
inhomogeneous sample). This is marginally significant ($2\sigma$
deviation) and implies a change in the physical nature and/or extent
of surface features when moving from M to L dwarfs.

}

\maketitle

\input{Introduction}

\input{DataAcquisition}

\input{DataReduction}

\input{TimeSeriesAnalysis}

\input{Results}

\input{Conclusions}

\begin{acknowledgements}
We are very grateful to Roland Gredel, the director of the Calar Alto
observatory, for a prompt and uncomplicated allocation of Directors
Discretionary Time in 2005 and to the Calar Alto staff for obtaining these
data and for their support during the 2002 and 2003 observing runs.
\end{acknowledgements}

\bibliographystyle{aa}
\bibliography{Bibliography}

\begin{longtable}{llllllllrll}
\caption{\label{table_TargetPropertiesTwo} Summary of important results:
  $N_{rs}$ is the number of reference stars, $\chi^{2}/\chi^{2}_{0}$ the
  actual and critical $\chi^{2}$ values (i.e. for $p=0.01$), var.\ flags
  indicates whether a target is variable on the individual nights and in total
  (the latter in parenthesis) according to the $\chi^{2}$ test, $p_{tot}$ is
  the probability of the null hypothesis, ampl.\ the variability amplitude
  (root-mean-square) or an upper limit hereto (both in magnitudes), period is
  self-explanatory (measured in hours with uncertainties obtained by the
  simulations, see Sect.~\ref{PeriodUncertainty}, although note that
  theoretical values are larger); AR (Amplitude Ratio) is the ratio of the rms
  amplitude of the light curve to the noise within the target; \emph{var.}
  states whether or not a channel was finally considered variable. The $+$ and
  $-$ flags indicate variable and non-variable objects, respectively, whereas
  o means that it is only a marginal detection (either in terms of the
  $\chi^{2}$ test or of a consideration of all means of investigation).}\\
\hline\hline
target  &  SpT  &  band  &  $N_{\rm rs}$  &  $\chi^{2}/\chi^{2}_{0}$  &
var.\ flags  &  $p_{\rm tot}$  &  ampl.\  &  period  &  AR  &  var.\\
\hline
\endfirsthead
\caption{continued.}\\
\hline\hline
target  &  SpT  &  band  &  $N_{\rm rs}$  &  $\chi^{2}/\chi^{2}_{0}$  &
var.\ flags  &  $p_{\rm tot}$  &  ampl.\  &  period  &  AR  &  var.\\
\hline
\endhead
\hline
\endfoot

2M1311+80   &  M8  &  I  &  $17$  &  38.3/54  &  --, o (--)  &  $0.21$  &  $0.012^{a}$  &  ---  &  $1.2$  &  --\\
            &      &  R  &  $12$  &  20.1/26  &  --   (--)  &  $0.065$  &  $0.020^{a}$  &  ---  &  $1.4$  &  --\\
            &      &  G  &  $13$  &  58.5/54  &  --, -- (+)  &  $2.9\cdot10^{-4}$  &  $0.088^{b}$  &  ---  &  $1.4$  &  --\\
2M1336+47   &  M7  &  I  &   $9$  &  72/95    &  --, -- (--)  &  $0.71$  &  $0.0090^{a}$  &  ---  &  $1.3$  &  --\\
            &      &  R  &  $10$  &  66.9/95  &  --, -- (--)  &  $0.41$  &  $0.022^{a}$  &  ---  &  $1.0$  &  --\\
            &      &  G  &   $9$  &  62.2/96  &  --, -- (--)  &  $0.61$  &  $0.063^{a}$  &  ---  &  $0.8$  &  --\\
2M1344+77   &  M7  &  I  &  $33$  &  63.5/64  &  --, + (o)  &  $0.011$  &  $0.0057^{a}$  &  ---  &  $2.0$  &  o\\
            &      &  R  &  $18$  &  51.8/58  &  --, -- (--)  &  $0.034$  &  $0.016^{a}$  &  ---  &  $1.1$  &  --\\
            &      &  G  &  $19$  &  28.7/64  &  --, -- (--)  &  $0.91$  &  $0.048^{a}$  &  $12.5\pm 3.0$ ?  &  $0.6$  &  --\\
LHS370      &  M5  &  I  &  $14$  &  57.7/37  &  +  &  $<10^{-6}$  &  $0.015^{b}$  &  ---  &  $2.4$  &  +\\
            &      &  R  &  $13$  &  19.7/34  &  --  &  $0.29$  &  $0.020^{a}$  &  $5.9\pm 2.0$  &  $1.2$  &  +\\
            &      &  G  &  $14$  &  16.1/37  &  --  &  $0.65$  &  $0.045^{a}$  &  $6.5\pm 2.0$  &  $1.1$  &  +\\
LHS2930     &  M6.5  &  I  &  $10$  &  80/85    &  +, -- (--)  &  $0.024$  &  $0.026^{a}$  &  $13.2\pm 1.9$ ?  &  $2.4$  &  o\\
            &        &  R  &  $16$  &  55/85    &  --, -- (--)  &  $0.53$  &  $0.0088^{a}$  &  ---  &  $2.1$  &  --\\
            &        &  G  &  $10$  &  56/85    &  --, -- (--)  &  $0.52$  &  $0.063^{a}$  &  ---  &  $1.1$  &  --\\
CTI1539+28  &  M2  &  I  &  $29$  &  74.9/75  &  --, o (o)  &  $0.01$  &  $0.0074^{a}$  &  ---  &  $2.8$  &  o\\
            &      &  R  &  $21$  &  40.9/75  &  --, -- (--)  &  $0.79$  &  $0.0063^{a}$  &  ---  &  $0.8$  &  --\\
            &      &  G  &  $15$  &  67.0/74  &  --, -- (--)  &  $0.036$  &  $0.014^{a}$  &  ---  &  $1.4$  &  --\\
2M1546+37   &  M7.5  &  I  &  $25$  &  58.7/75  &  --, --, -- (--)  &  $0.84$  &  $0.0063^{a}$  &  ---  &  $1.2$  &  --\\
            &        &  R  &  $23$  &  105/75   &  +, +, -- (+)  &  $5\cdot 10^{-6}$  &  $0.012^{b}$  &  ---  &  $2.7$  &  +\\
            &        &  G  &  $18$  &  75/75    &  --, --, -- (o)  &  $0.01$  &  $0.025^{a}$  &  ---  &  $1.8$  &  --\\
LHS3189     &  M5.5  &  I  &  $35$  &  59.3/74  &  o, -- (--)  &  $0.13$  &  $0.0056^{a}$  &  ---  &  $1.1$  &  --\\
            &        &  R  &  $27$  &  35.6/74  &  --, -- (--)  &  $0.91$  &  $0.0097^{a}$  &  ---  &  $0.7$  &  --\\
            &        &  G  &  $18$  &  22.2/74  &  --, -- (--)  &  $0.9995$  &  $0.032^{a}$  &  ---  &  $0.6$  &  --\\
2M1627+81   &  M9  &  I  &  $36$  &  56.8/78  &  --, --, -- (--)  &  $0.37$  &  $0.011^{a}$  &  ---  &  $1.0$  &  --\\
            &      &  R  &  $31$  &  74.0/78  &  --, --, + (--)  &  $0.019$  &  $0.024^{a}$  &  ---  &  $1.4$  &  --\\
            &      &  G  &  $37$  &  41.3/78  &  --, --, -- (--)  &  $0.83$  &  $0.051^{a}$  &  ---  &  $0.8$  &  --\\
CTI1629+28  &  M4  &  I  &  $28$  &  44.1/74  &  --, -- (--)  &  $0.64$  &  $0.012^{a}$  &  ---  &  $1.0$  &  --\\
            &      &  R  &  $19$  &  206/74   &  +, + (+)  &  $<10^{-6}$  &  $0.014^{b}$  &  ---  &  $1.9$  &  +\\
            &      &  G  &  $20$  &  86.8/73  &  +, -- (+)  &  $4\cdot 10^{-4}$  &  $0.022^{b}$  &  ---  &  $1.5$  &  --\\
2M1707+64   &  M9  &  I  &  $37$  &  143/54   &  +, + (+)  &  $<10^{-6}$  &  $0.012^{b}$  &  $3.65\pm 0.1$  &  $2.5$  &  +\\
            &      &  R  &  $32$  &  62/54    &  +, + (+)  &  $0.001$  &  $0.014^{b}$  &  $3.7\pm 0.1$  &  $1.7$  &  +\\
            &      &  G  &  $25$  &  50/54    &  --, + (--)  &  $0.022$  &  $0.029^{a}$  &  $3.3\pm 0.1$  &  $1.8$  &  +\\
CTI1709+27  &  M5.5  &  I  &  $43$  &  120/80  &  --, --, + (+)  &  $<10^{-6}$  &  $0.0062^{b}$  &  ---  &  $2.1$  &  +\\
            &        &  R  &  $26$  &  277/80  &  --, +, + (+)  &  $<10^{-6}$  &  $0.011^{b}$  &  ---  &  $2.2$  &  +\\
            &        &  G  &  $29$  &  140/78  &  +, +, + (+)  &  $<10^{-6}$  &  $0.014^{b}$  &  ---  &  $1.6$  &  +\\
2M1714+30   &  M6.5  &  I  &  $32$  &  30.0/77  &  --, -- (--)  &  $0.99$  &  $0.0080^{a}$  &  ---  &  $0.8$  &  --\\
            &        &  R  &  $31$  &  107/77  &  +, + (+)  &  $<10^{-6}$  &  $0.012^{b}$  &  $6.9\pm 0.15$ ?  &  $1.9$  &  +\\
            &        &  G  &  $27$  &  144/77  &  +, + (+)  &  $5\cdot 10^{-6}$  &  $0.034^{b}$  &  ---  &  $1.8$  &  --\\
LHS3307     &  M5  &  I  &  $49$  &  67.5/84  &  --, -- (--)  &  $0.13$  &  $0.052^{a}$  &  ---  &  $>3$  &  --\\
            &      &  R  &  $50$  &  104/84  &  +, -- (+)  &  $0.0001$  &  $0.012^{b}$  &  ---  &  $1.2$  &  o\\
            &      &  G  &  $47$  &  57.7/84  &  --, -- (--)  &  $0.41$  &  $0.031^{a}$  &  ---  &  $0.8$  &  --\\
2M1750+44   &  M7.5  &  I  &  $31$  &  84.3/92  &  --, + (--)  &  $0.038$  &  $0.0075^{a}$  &  ---  &  $1.1$  &  --\\
            &        &  R  &  $36$  &  62.4/88  &  --, -- (--)  &  $0.36$  &  $0.018^{a}$  &  ---  &  $1.3$  &  --\\
            &        &  G  &  $33$  &  69.7/80  &  o, -- (--)  &  $0.061$  &  $0.036^{a}$  &  ---  &  $1.2$  &  --\\
LHS3339     &  M6  &  I  &  $32$  &  116/75  &  --, -- (+)  &  $<10^{-6}$  &  $0.0061^{b}$  &  ---  &  $1.7$  &  +\\
            &      &  R  &  $31$  &  67.6/74  &  --, -- (--)  &  $0.033$  &  $0.0083^{a}$  &  ---  &  $1.5$  &  --\\
            &      &  G  &  $20$  &  119/75  &  --, -- (+)  &  $<10^{-6}$  &  $0.021^{b}$  &  ---  &  $1.3$  &  +\\
2M1757+70   &  M7.5  &  I  &  $50$  &  48.6/60  &  --, -- (--)  &  $0.097$  &  $0.0061^{a}$  &  ---  &  $4.7$  &  --\\
            &        &  R  &  $39$  &  54.8/60  &  --, -- (--)  &  $0.03$  &  $0.0090^{a}$  &  ---  &  $1.5$  &  --\\
            &        &  G  &  $29$  &  50.7/60  &  --, + (--)  &  $0.066$  &  $0.025^{a}$  &  ---  &  $1.8$  &  --\\
CTI1801+28  &  M2.5  &  I  &  $31$  &  49.0/84  &  --, -- (--)  &  $0.73$  &  $0.0061^{a}$  &  ---  &  $1.1$  &  --\\
            &        &  R  &  $31$  &  73.7/81  &  --, -- (--)  &  $0.039$  &  $0.0055^{a}$  &  ---  &  $1.2$  &  --\\
            &        &  G  &  $27$  &  63.2/81  &  --, -- (--)  &  $0.18$  &  $0.015^{a}$  &  ---  &  $1.1$  &  --\\
LHS3376     &  M4  &  I  &  $44$  &  135/50  &  +  &  $<10^{-6}$  &  $0.052^{b}$  &  ---  &  $>3$  &  --\\
            &      &  R  &  $25$  &  27/50  &  --  &  $0.43$  &  $0.029^{a}$  &  ---  &  $1.3$  &  --\\
            &      &  G  &  $40$  &  45/50  &  --  &  $0.029$  &  $0.050^{a}$  &  ---  &  $1.9$  &  --\\
\hline
\multicolumn{8}{l}{amplitude column:   $^{a}$ upper-limit-rms;    $^{b}$ light curve rms}\\
\end{longtable}

\end{document}

%% file: Introduction.tex
\section{Introduction}
Photometric and spectroscopic monitoring of very low mass stars and
substellar objects has been performed over the last ten years by
several groups and has led to new insights into the processes
occurring in these objects. Since stars on the main sequence become
fully convective below a mass of about $0.3-0.4\,M_{\odot}$ (roughly
corresponding to a spectral type of M3 to M4), magnetic fields cannot
be maintained in stars of lower mass or later type via an
$\alpha\Omega$ dynamo. 
An alternative dynamo may come into operation, for example
an $\alpha^{2}$-dynamo (Chabrier \& K\"uker astro-ph/0510075).
A change in the dynamo mechanism with spectral type could
directly affect the activity of M~dwarfs as measured by H$_{\alpha}$
emission, for example. Early M dwarfs in most cases do not show any
detectable H$_{\alpha}$ emission whereas it increases at mid and late
M type. Furthermore, a saturation-type relation between rotation and
activity has been confirmed by \citet{Mohanty2003} and
\citet{Delfosse1998}. The detection of flares on these objects also
proves the existence of magnetic fields. However, magnetic activity
strongly decreases again with early L type \citep{West2004} which can
be explained via an increasingly neutral photosphere and the onset of dust
formation \citep{Mohanty2002}. Since photometric variability can be
caused by both magnetic activity (through star spots) and dust clouds,
it is not clear whether variability is more frequent or less frequent
in M dwarfs than in L dwarfs. Measurements of $v\sin i$ values show
that later-type objects on average rotate much faster and lead to
expected rotation periods of about $1.5-13\,{\rm hr}$
\citep{Bailer-Jones2004} for L dwarfs, whereas M dwarfs can have
significantly longer periods of up to two days. Thus it appears that
the spin-down timescale is much longer for later-type objects.

Stellar/substellar rotation periods can be measured directly from
photometric time series if the objects show surface brightness
inhomogeneities.  This also permits a study of multiperiodicity and/or
nonperiodic variability. Candidates for the surface features are
magnetically induced spots and, because of their lower effective
temperature, dust clouds for late~M and L~dwarfs. By modelling the
variability at different wavelengths, we can attempt to infer the
physical properties of the atmosphere and surface features in
individual cases. The few attempts to attribute a specific physical
cause of the variability (via time-resolved spectroscopy) have been
rather inconclusive
\citep{Bailer-Jones2002,Clarke2003,Bailer-Jones2003}, however, partly
due to the high multifrequency photometric sensitivity (1\,\% or
better) required to distinguish between mechanisms.

There have been several one- or two-band photometric surveys of ultra
cool dwarfs (UCDs) of both young cluster objects and older field
objects, mostly in the I-band but also in the R or near infrared
bands. A comparison of these reveals variability (periodic or
nonperiodic) to be present in about $40\,\%$ of objects in each of
these two age groups \citep{Bailer-Jones2005}. Interestingly,
periodic variability is more common among the younger cluster sample,
whereas cases of nonperiodic variability are more frequent in the
older field sample. This could, in principle, be either an age or a
spectral type effect. An age effect might be related to a decline in
activity, possibly from the dissipation of disks, although we have no explicit
mechanism for this. A spectral type effect
could be related to the onset of dust formation and, specifically, the
dynamics of large scale dust clouds: \citet[ henceforth
BJM01]{Bailer-Jones2001} proposed a ``masking hypothesis'' in which
the rapid formation and evolution of dust clouds masks the rotational
modulation of the light curves of L dwarfs, thus accounting for their
nonperiodic variability (also see \citet{Bailer-Jones2004}). We
cannot distinguish between these two effects in the samples to date,
because there is a broad age--spectral type (SpT) correlation in the
object selection: most M~dwarfs monitored were in young clusters
(1--100 Myr) whereas the L dwarfs were field objects (probably more
than several hundred Myr in age). Specifically, in our earlier work
(BJM01) -- where we monitored 21 M6--L5 dwarfs -- we found variablity
to be more common for objects later than M9. However, as stated in
that paper, because of a coarse age--SpT correlation in the sample, we
could not distinguish between an age and a SpT dependence based only
on those data.

The main objective of the present work is to remove this correlation
by extending the survey of BJM01 to (older) field M dwarfs.  This
represents a control sample against which the occurrence, periods,
amplitudes and nature of variability in the field L dwarf sample of
BJM01 can be compared, i.e.\ two populations with similar ages.  Early
and mid M dwarfs do not show photospheric dust formation (either
theoretically or observationally), so together with the L dwarf
samples allows us to study the link between variability and dust. In
addition to significantly increasing the time-resolved data on field M
dwarfs in G, R and I, we are also able to constrain the physical cause
of the variability in individual cases.

The structure of the paper is as follows: The next section describes
the target selection and data acquisition process. In
Section~\ref{DataReduction} we present the basic data reduction steps,
photometry and a reliable error description model which together allow
us to investigate low-amplitude variability. Methods for analysing
general and periodic variability are discussed in
Section~\ref{TimeSeriesAnalysis}; here we also describe our Monte
Carlo approach to estimating the reliability and uncertainty in the
period detection process. We briefly mention the multichannel
detection of a massive flare in an M9 dwarf. This is reported in more
detail in \citet{Rockenfeller2006}. Section~\ref{Results} deals with
the overall results plus details on some individual targets. Two of
these objects are studied for the cause of the observed variability
and a comparison of the frequency of variability for L and M dwarfs is
performed. We finish with the conclusions in
Section~\ref{Conclusions}. For a more detailed description and
discussion of the methods and for comments on every individual target
see \citet{Rockenfeller2005a}.


%% file: DataAcquisition.tex
\section{Data Acquisition}

\subsection{Target selection}
The target list was assembled by selecting bright M~dwarfs ($m_{I}<17.0\,{\rm
mag}$) which are visible from the Calar Alto observatory, Spain, in June for
several hours at an airmass of less than~$2.0$. 
Many of the objects are bright enough such that 
the G-band data have a relatively high signal-to-noise ratio
(SNR). Strong activity as measured by H$_{\alpha}$ was avoided and they were
chosen to cover most of the spectral M-type sequence (M2 to M9). A total of 19
field M-dwarfs were observed over the two observing seasons. Target details
are given in Table~\ref{table_TargetPropertiesOne}.
\begin{table*}
\caption{Target properties: full name, spectral type, I magnitude,
  number of images in total, year of observation, nights of observations,
  duration of the observations on the individual nights, additional
  information and the references to these. Objects are sorted by increasing
  right-ascension.}
\label{table_TargetPropertiesOne}
\centering
\begin{tabular}{lllllllll}
\hline
\hline
target  &  SpT  &  $m_{I}$  &  $N_{obs}$  &  year  &  nights  &  duration [hr]
&  further information  &  reference\\
\hline
\object{2MASSW J1311391+803222}   &  M8    &  15.9  &  33  &  2002  &  5,8  &  5.9, 6.7
&  $H_{\alpha}=3.0$  &  G00\\
\object{2MASSW J1336504+475131}   &  M7    &  15    &  66  &  2003  &  9,10  &  4.6,
4.8  &  $v\cdot\sin i=30$, $H_{\alpha}=3.2, 5.0$  &  G00, R02\\
\object{2MASSW J1344582+771551}   &  M7    &  15.2  &  41  &  2002  &  1,4  &  5.2, 6.2  &  $H_{\alpha}=2.7$  &  G00\\
\object{LHS370}      &  M5    &  12.3  &  20  &  2002  &  8  &  6.1  &  -  &  -\\
\object{LHS2930}     &  M6.5  &  13.3  &  58  &  2003  &  7,8  &  7.1, 7.0  &
$R/R_{Sun}\approx 0.33$, $T_{\rm eff}= 2687\pm 65\,$K  &  H04, D02\\
\object{CTI153948.1+280322}  &  M2    &  14.9  &  50  &  2003  &  1,3  &  4.9, 6.9  &  Binary Star; $v\,\sin i=7.0$,
$H_{\alpha}=0.7$  &  M03\\
\object{2MASSW J1546054+374946}   &  M7.5  &  15.2  &  51  &  2003  &  4,5,6  &  6.4, 2.6, 2.4  &
 $H_{\alpha}=10.9$  &  G00, C03\\
\object{LHS3189}     &  M5.5  &  14.9  &  49  &  2002  &  1,2  &  6.5, 6.7  &  -  &  B91\\
\object{2MASS J16272794+8105075}   &  M9    &  16.4  &  52  &  2003  &  4,5,6  &  6.4, 2.9, 2.4  &  &  G00\\
\object{CTI162920.5+280239}  &  M4    &  15.6  &  49  &  2003  &  1,3  &  5.2, 6.4  &  $H_{\alpha}=10.9$  &  -\\
\object{2MASSW J1707183+643933}   &  M9    &  15.9  &  34  &  2002  &  4,5  &  6.4, 5.6  &  $T_{\rm eff}\approx 2350\,$K,
$H_{\alpha}=9.8$,  &\\
            &        &        &      &        &       &            &  $\log g=5.16$,
$M/M_{Sun}=0.062$  &  G00, G03\\
\object{CTI170958.5+275905}  &  M5.5  &  13.9  &  54  &  2003  &  4,5,6  &  6.7, 2.8,
2.3  &  &  K94\\
\object{2MASSW J1714523+301941}   &  M6.5  &  14.9  &  51  &  2003  &  1,3  &  6.1, 5.4  &  $T_{\rm eff}\approx 2775\,$K, $v\cdot\sin i<4$,  &\\
            &        &        &      &        &       &            &
            $H_{\alpha}=3.2, 5.4$  &  G00, R02\\
\object{LHS3307}     &  M5    &  15     &  57  &  2003  &  9,10  &  6.8, 4.1  &  -  &  B91\\
\object{2MASS J17501291+4424043}   &  M7.5  &  15.6  &  64  &  2003  &  7,8  &  6.8, 6.9
&  $H_{\alpha}=2.7$  &  G00\\
\object{LHS3339}     &  M6    &  14.0  &  50  &  2002  &  1,2  &  6.0, 6.4  &  $T_{\rm
  eff}= 2957\pm 70\,$K,
$R/R_{Sun}\approx 0.33$  &  L00, D02\\
\object{2MASSW J1757154+704201}   &  M7.5  &  14.2  &  38  &  2002  &  4,5  &  6.3, 7.0
&  $H_{\alpha}=3.0$  &  G00\\
\object{CTI180120.1+280410}  &  M2.5  &  14.8  &  57  &  2003  &  7,8  &  6.2, 6.9  &  -  &  -\\
\object{LHS3376}     &  M4    &  10.7  &  30  &  2002  &  8  &  6.0  &  $T_{\rm eff}\approx 3100\,$K,
$v\cdot\sin i=14.6\pm 1.0$  &  D98\\
\hline
\multicolumn{8}{l}{B91: \citet{Bessell1991}; K94: \citet{Kirkpatrick1994}; D98:
  \citet{Delfosse1998}; G00: \citet{Gizis2000}}\\
\multicolumn{8}{l}{L00: \citet{Leggett2000}; R02:
  \citet{Reid2002}; D02: \citet{Dahn2002}}\\
\multicolumn{8}{l}{M03: \citet{Mohanty2003}; C03: \citet{Cruz2003}; G03:
  \citet{Gorlova2003}; H04: \citet{Henry2004}}\\
\end{tabular}
\end{table*}

\subsection{Observations}
\begin{figure}
  \resizebox{\hsize}{!}{\includegraphics{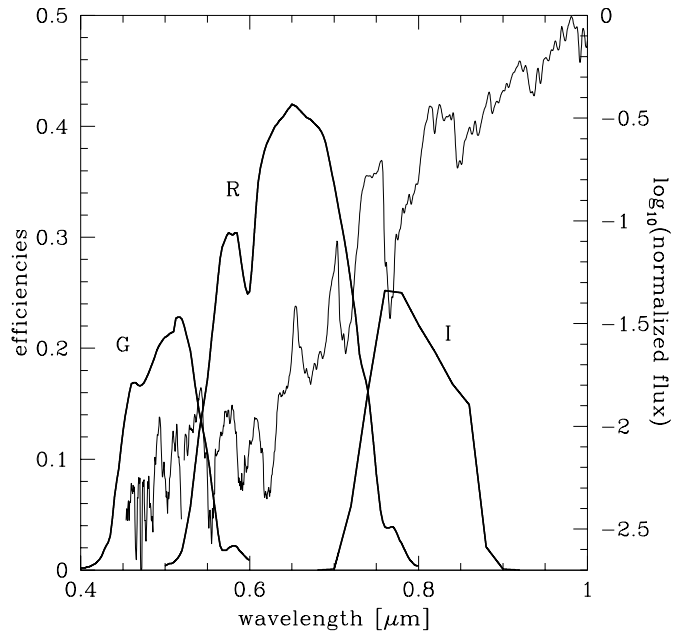}}
  \caption{The total efficiencies of the BUSCA G, R and I channels,
    taking into account the dichroics, the CCD efficency and the
    Bessell-I filter used in the I-band. Also plotted is the spectrum of
    the M4.5~dwarf DENIS$\,$P-J1158-1201 of \citet{Martin1999}.} 
  \label{figure_BUSCAwavebands}
\end{figure}

To perform simultaneous multiband photometry, we used the BUSCA four
channel CCD camera at the $2.2\,$m telescope at Calar Alto Observatory,
Spain. This instrument uses dichroics to split the light beam into
four wavelength bands, namely UV, G, R and I. The UV, G and R passbands are
defined by the CCD response and dichroic transmission function; for
the I band we additionally used a Bessel I filter (see
Fig.~\ref{figure_BUSCAwavebands}).
Each channel is equipped with a CCD485 Lockheed Martin 4Kx4K CCD.  The
CCDs are thick, except for the UV channel which uses a
backside-illuminated thinned chip. For all CCDs 2x2 binning was used,
reducing read-out time and leading to a pixel scale of $0.352\arcsec$. The
field-of-view of $12\arcmin\cdot 12\arcmin$ allowed us to select a
large number of reference stars for differential photometry. For use
throughout this paper, we define an adjusted Julian day (AJD) by ${\rm
AJD}={\rm JD}-2\,450\,000$. Observations were conducted in two runs:
the first with 8 usable nights in June 2002 (AJD 2424.4--2431.7) and
the second with 9 usable nights in June 2003 (AJD
2794.4--2803.6). On each night, two or three targets were observed
alternately and for each target, data were collected on two or three
nights. Integration times of $45$ to $300\,$s were used depending on
target brightness and weather conditions, to achieve high
signal-to-noise data and still retain sensitivity to periods in the
sub-hour regime.


%% file: DataReduction.tex
\section{Data reduction}\label{DataReduction}

\subsection{Basic reduction steps}
All science images were reduced using the IRAF\footnote{The Image
Reduction and Analysis Facility, provided by the National Optical
Astronomy Observatories (NOAO).} package. UV-band data could not be
used due to the faintness of the M dwarfs in this spectral range,
other than for finding flares in the targets' time series (see
Section~\ref{Flares}).  We reduced the other three channels with the
same reduction steps, with parameters set to achieve homogeneous,
comparable photometry. The basic reduction steps include an overscan
subtraction and flat fielding process. Because non-thinned CCDs were
used, no fringes can be seen in any of the images and hence fringe
correction was not performed.

The zero integration time frames, which were obtained to correct for
two-dimensional bias patterns in the science images, are useless: They
all show vertical bars caused by the read-out electronics which do not
appear in the science frames, a known problem that occurs frequently
when using BUSCA. To partly compensate for this, a first-order fit to
the overscan area was subtracted from each science and flat field
frame. This corrects at least for one-dimensional patterns in the
y-direction of the bias level. Because residual two-dimensional
patterns are only weak and do not show a strong gradient on the scale
of the applied sky apertures, their contribution to the total
photometric error is well below $0.5\,\%$. We used an clipped average
of ten dome flat fields to correct for changes in the pixels' quantum
efficiencies. Individual flat field frames were applied separately for
each channel and each night.

\subsection{Photometry}
To achieve the desired accuracy of better than $0.01\,$mag,
differential photometry was performed to reduce the effects of
temporal variations in Earth's atmosphere. Aperture photometry was
done on the target and tens of reference stars; these were chosen
according to the criteria used in BJM01. Refer also to that paper for a
description of differential photometry. 
Although various aperture radii were tested, the results
we show are with a radius of 6 pixels (except for LHS370, see
Section~\ref{Results_Individual}) which we found to maximize the
signal-to-noise ratio (SNR). The large field-of-view guarantees that
in most cases more than $15$ (and even up to $50$) non-variable
reference stars could be found within the target field, ensuring a
stable reference flux. For the procedure of creating the reference
star list, see the next section.
We excluded the presence of significant second-order-extinction (SOE;
see~\citet{Bailer-Jones2003} for a discussion) in our data by:
plotting relative magnitudes versus airmass; comparing the relative
light curves formed with only blue and only red reference stars;
plotting the variability measure of all reference stars of a field
($\chi^{2}$ values, see Section~\ref{ChisqTest}) versus the $R-I$
colour. See \citet{Rockenfeller2005a} for more details.

\subsection{Error sources and estimation}
An accurate error estimate is important for assessing the presence of
variability. As discussed in \citet{Bailer-Jones1999}, the total photometric
error is composed of \emph{formal} (Poisson noise) and \emph{informal} error
contributions. As imperfect flat fielding is thought to dominate the
informal error, we first tried to model the total photometric error
($\sigma_{\rm tot}$) by adding a constant term (of $0.5\,\%$) in
quadrature to the theoretical (formal) errors provided by IRAF
($\sigma_{\rm IRAF}$). However, this lead to an undesirable magnitude
dependence of the variability measure $\chi^{2}$. Instead (and to
overcome this), we finally used the following error model: because all
reference stars of a target field are non-variable, the scatter
(standard deviation $\sigma_{\rm rms}$) in their light curves is a
measure of the total photometric error at the corresponding
magnitude. We plotted $\sigma_{\rm rms}$ versus $\sigma_{\rm IRAF}$
and fitted a first order polynomial to this graph. Applying this fit,
i.e. $\sigma_{\rm tot}=a+b\cdot\sigma_{\rm IRAF}$ where $a$ and $b$
are the fit's free parameters, we arrive at a reliable error
description. Typical values of $a$ and $b$ are $-0.01$ to $0.01$ and
$0.5$ to $1.5$, respectively. The fitting procedure was done for each
field and channel separately. The reference star selection is an
iterative process, which means that first the $\chi^{2}$ measure (see
Section~\ref{ChisqTest}) is determined for all candidates. Then the
most variable one is excluded from the set and the process is repeated
until no variable star remains. Our error model partly accounts for
varying data quality on different nights, e.g.\ due to bad seeing
conditions, since this manifests itself as an increased standard
deviation in the reference stars.

We further found that other informal errors contribute far less than
$0.5\,\%$. This includes imperfect overscan subtraction and two rather unusual
sources: weak charge trailing on the CCD for brighter stars (and which
appears only in the G-channel) and what we call \emph{glitches}. These
are quite similar in size and intensity to cosmics but are caused by
the read-out electronics. They are quite numerous in the I-channel
(some hundred per image) but only rare in R and G.


%% file: TimeSeriesAnalysis.tex
\section{Time series analysis}\label{TimeSeriesAnalysis}

\subsection{$\chi^{2}$ test}\label{ChisqTest}
To test whether or not deviations in a target's relative light curve are
consistent with the photometric errors (the null hypothesis) we evaluated the
$\chi^{2}$ measure
\begin{displaymath}
\chi^{2}=\sum^{N}_{i}\left(\frac{m_{\rm rel}(i)}{\delta m_{\rm rel}(i)}\right)^{2},
\end{displaymath}
where $m_{\rm rel}(i)$ is the relative magnitude in the $i$th (of $N$) frame
and $\delta m_{\rm rel}(i)$ the error therein. The larger the $\chi^{2}$ value,
the larger the probability that the null hypothesis is wrong and that
the object is variable. We claim an object to be variable if the probability
for the null hypothesis $p$ is smaller than $0.01$. We used this test first to
assemble the set of reference stars and then to test for possible variability
within the targets.

We often encountered the case that significant general or periodic variability
is only present in one or two channels. This can either be due to different
sensitivity limits or to different variability amplitudes in the channels (or
to a combination of both). As we will discuss in chapter~\ref{Results},
surface features can indeed lead to such a behaviour.

\subsection{Periodic variability}\label{PeriodicVariability}
Periodic variability can be caused by co-rotating surface features which are
stable on time scales of an object's rotation period. To check for periodic
behaviour, the following scheme was applied to each target: the CLEAN
periodogram was calculated and searched for significant peaks; in case there
is such a peak, the target's light curve was phased to the corresponding
period and checked if it confirms the period in question. The Lomb-Scargle
periodogram was also evaluated for all targets but the influence of the
spectral window function is very strong for periods longer than approximately
$10$ hours. Because of this the Scargle power spectra are not used for the
final results. Nevertheless, the shorter periods of 2M1707+64 and LHS370 are
confirmed by this method.

The CLEAN algorithm tries to remove the influence of the discrete and finite
sampling of observational data on power spectra. For more information on this,
see BJM01 and \citet{Roberts1987}. To judge whether a peak in the power spectrum
is statistically significant, we performed Monte-Carlo simulations to determine
the \emph{false alarm probability} (FAP) power levels for each time
sampling. The FAP denotes the probability that a peak in the power
spectrum is caused by noise. We chose to use the same methods as already
discussed by \citet{Lamm2004}. Artificial light curves of non-variable stars
were created by simulating pure Gaussian noise as well as by shuffling the
actual magnitudes of the target's light curve with respect to the real
epochs. For both methods, the highest peak in the corresponding CLEAN
periodogram was determined and the power level that was exceeded by $100$ out
of the $10^{4}$ simulations was defined to be the $1\,\%$ FAP power level
(similarly one obtains the $10\,\%$ and $0.1\,\%$ levels). To be conservative
we compare the corresponding two values obtained for both methods and use the
larger one. We claim periodic variability if peaks above the $1\,\%$ limit are
present.

\subsubsection{Period uncertainty}\label{PeriodUncertainty}
\begin{figure}
  \resizebox{\hsize}{!}{\includegraphics{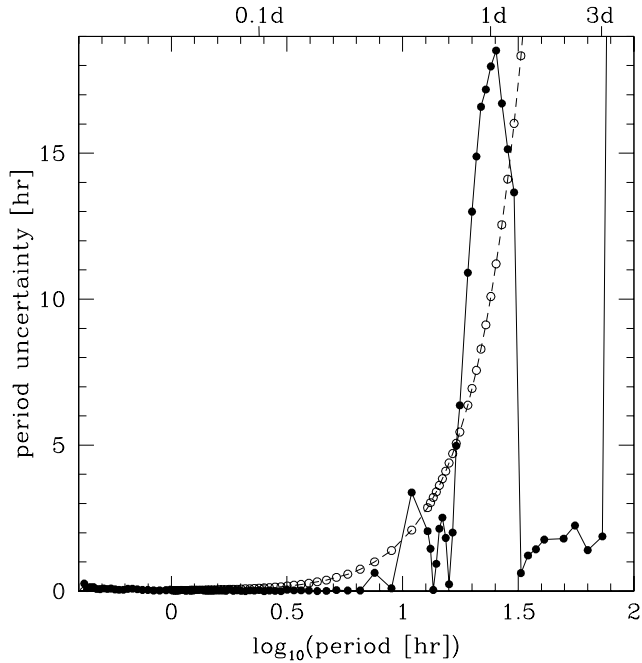}}
  \caption{Period uncertainty in hours plotted versus the logarithm of
  the period in hours; obtained for an amplitude ratio of $1.5$ for the
  2M1707+64 I time sampling with the method described in
  section~\ref{PeriodicVariability} (solid line and solid circles). The
  results of equation~\ref{equation_PeriodUncertainty} are also included as
  open circles on a dashed line.}
  \label{figure_PeriodUncertainty}
\end{figure}

To investigate the uncertainty in the detected periods, we simulated a
sinusoidal signal at random phase with respect to the time sampling and
added Gaussian noise. We define the Amplitude Ratio (AR) to be the ratio
between the root-mea-square (rms) amplitude of the sinusoidal and the one of
the noise. The absolute value of the difference between the period of the input
signal and that of the highest peak in the CLEAN power spectrum was averaged
over $10^{4}$ simulations at each input period. This result is a measure for
the expected period uncertainty at the period of the sinusoid. An example for
one time sampling is shown in Fig.~\ref{figure_PeriodUncertainty}. The input
frequency is varied over the whole range of frequencies (actually we divided
that range into $100$ bins and performed $10^{4}$ simulations per bin).
The shape of the curve is only weakly dependent on the AR for values
above $1.5$ and period uncertainties only slightly increase for ARs down to
$1.0$. Table~\ref{table_TargetPropertiesTwo} shows that, except for one case,
all claimed periods occur at ARs of more than one. The simulations
show that the period accuracy is very good for periods up to about $10$ hours
and becomes much worse between $15$ and $30$ hours. This can be attributed to
the gap in the data between the two consecutive observation nights. Although
we expect the uncertainty to become lower for yet longer periods (since we
observed the target on two nights), it is surprising how low it actually
gets. A possible problem of these simulations is that they assume a sinusoidal
signal, whereas periodic variability does not necessarily match this.

A similar procedure to the above has been used by
\citet{Scholz2004a,Scholz2004b} who find a similar behaviour of the period
uncertainty. Also shown in Fig.~\ref{figure_PeriodUncertainty} (open circles)
are the results of a theoretical period uncertainty estimate (for small
errors): 
\begin{equation}\label{equation_PeriodUncertainty}
\Delta P\approx\frac{\Delta\nu P^{2}}{2}.
\end{equation}
Here $\Delta\nu$ is the width of the main peak of the window function
$W(\nu)$, that can be approximated for not too uneven data sampling by
$\Delta\nu\approx\frac{1}{T}$. Here, $T$ is the total time span covered by the
observations. $P$ and $\Delta P$ are the period in question and its
uncertainty, respectively. For long periods (larger errors) the approximation
made to derive the above equation may no longer be valid. This would explain
the huge difference relative to the results obtained with the simulations at
longer periods. For more information on
equation~\ref{equation_PeriodUncertainty}, see \citet{Roberts1987}.

Since the predicted uncertainties may differ by a factor of up to 3 between
these two methods and since both methods do have drawbacks, it is difficult to
decide which one to use. Thus the period uncertainties we state are always the
results of the simulations but we also give the values obtained by
equation~\ref{equation_PeriodUncertainty} in parenthesis. Apart from just
finding errors on the periods found with this survey, the period uncertainty
is important when periods are reported in more than one channel of a
target. If those periods are close to each other, it is likely that they
actually correspond to a single periodicity, e.g. the rotation period, if
their values do not differ by more than two or three times their uncertainty.

\subsubsection{Detection Fraction}\label{DetectionFraction}
\begin{figure}
  \resizebox{\hsize}{!}{\includegraphics{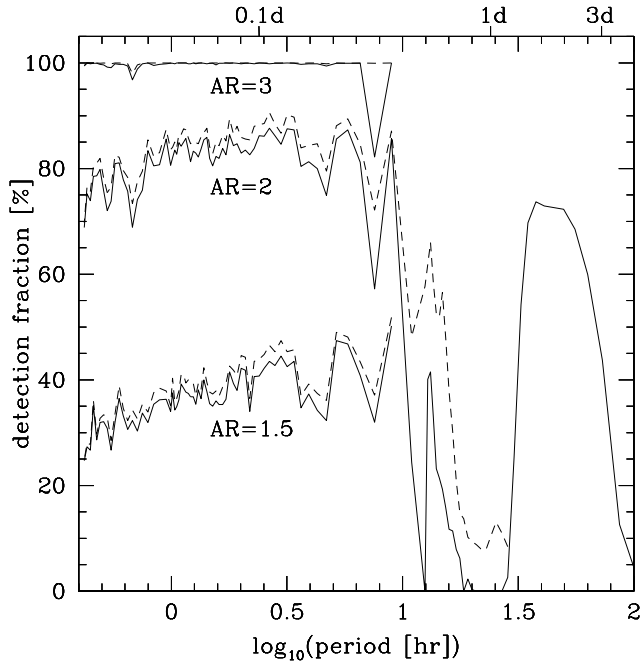}}
  \caption{The plot shows various detection fractions for the 2M1707+64 time
  sampling over the logarithm of the period in hours. Solid lines represent
  detection fractions which require a peak to be found in a narrow range
  around the input period (i.e. inside the input bin), (${\rm
  DF}_{\rm bin}$), and dashed lines the ones that allow a peak to be at any
  period (${\rm DF}_{\rm all}$). For periods of up to
  $\log_{10}($period$)=1.0$ three amplitude ratios are investigated: 3, 2, 1.5
  from top to bottom within the plot; for each ratio one solid and one dashed
  line. Because the general behaviour does not change, and to avoid crowding,
  for periods longer than ten hours only the curves corresponding to an
  amplitude ratio of 2 are shown. Here the solid and dashed lines merge at
  $\log_{10}($period$)=1.47$.}
  \label{figure_DetectionFraction}
\end{figure}

To study the sensitivity of our period detection procedure, we introduce the
\emph{detection fraction} (DF). This quantity is equal to the fraction of detected 
periodic signals as a function of period; it is calculated using simulations
of a sinusoidal and Gaussian noise (with the same definition of the AR as in
the last section). The whole range of frequencies was divided in $100$ bins
and the simulations performed separately for each bin. The fraction of
simulations with a fixed input frequency which lead to a peak in the power
spectrum above the $1\,\%$ FAP power level is equal to the DF. We applied two
different DFs, one where peaks anywhere in the periodogram were considered
(${\rm DF}_{\rm all}$) and another where a significant peak was counted only if
it was found inside the input bin (${\rm DF}_{\rm bin}$). Naturally, the
detection fraction depends strongly on the AR. See
Fig.~\ref{figure_DetectionFraction} for the results on the same target as in
Fig.~\ref{figure_PeriodUncertainty}.
The resemblance to the plot of the period uncertainty is striking and naturally
because both quantities depend on the same data sampling. Thus the same
reasons can be given for the minimum in the DF at about $P=10-30$ hr and
for the maximum in the period uncertainty at the same period range. The
difference between the solid and the dashed line in
Fig.~\ref{figure_DetectionFraction} is the fraction of period detections at
the wrong frequencies, i.e. outside the input bin. In agreement with the
period uncertainty simulations, this difference is quite small at periods
shorter than $10$ hours and thus if we detect a periodicity within this range
it is very likely detected at the right period. If we again compare the ARs
given in Table~\ref{table_TargetPropertiesTwo} with the Detection Fraction
plots of the individual targets, we can summarise the results as follows: for
most time samplings, an AR of $1.5$ leads to a DF between $50\,\%$ and
$80\,\%$; for ARs of $1.0$, the DF lies between $15\,\%$ and $30\,\%$ (in the
case of 2M1707+64, see Fig.~\ref{figure_DetectionFraction}, the DFs are
somewhat lower at each AR). Hence it is not unlikely that we missed periods
for targets with ARs of less than $1.5$, even in the range of up to ten
hours. Furthermore, the $v\sin i$ distribution of M dwarfs
\citep{Mohanty2003,Delfosse1998} suggests that there are some objects,
particularly at early M type, with expected periods of much more than $10$
hours (up to a few days). For these, the sensitivity of our detection
procedure is questionable. Hence, to lower the probability of missing periods
significantly, one would have to perform higher SNR observations and/or get a
longer time base.

\subsection{Flares}\label{Flares}
We will very briefly address the topic of flares in the observed field M
dwarfs. For detailed information on the flares detected in this data set, see
\citet{Rockenfeller2006}.

Although well-known, the exact processes taking place during a flare event are
not well-understood. Probably, magnetic energy is transfered to thermal energy
and thus leads to a brightening of the effected area on or near the stellar
surface. Flares on the Sun are often associated with eruptive promininences,
sometimes ejecting solar material and charged particles into the solar system.

In principle, multiband monitoring in optical bands is ideal to detect flares
because the amplitude of the brightness variations increases tremendously from
the I- to the UV-band. A flare is characterised by a fast rising signal
followed by a slower decreasing one (of exponential shape). The duration of
the event is positively correlated with its amplitude and ranges from a few
minutes (or even shorter) to a few hours. This creates the problem that
low-amplitude flares are too short to be seen in detail in this data set since
the minimum time span between two data points is about $5$ minutes. However,
one huge flare was detected in 2M1707+64, with an UV-amplitude of more than
$6$ magnitudes and a recorded duration of about $1$ hour. For this event we
captured the brightness evolution over five data points. Three other events
that are probably flares were found in 2M1714+30, 2M1546+37 and 2M1344+77 at
lower amplitudes (between $1.5$ and $2.7$ mag in UV). The total observation
time of this survey is $218.0$ hours, yielding a flare rate of $9.2\cdot
10^{-4}\,{\rm hr}^{-1}$, $1.83\cdot 10^{-3}\,{\rm hr}^{-1}$ if we only count
the two strongest events or all four, respectively.


%% file: Results.tex
\section{Results}\label{Results}

\subsection{General results}
The results of the $\chi^{2}$ test and the period search for all 19 targets
are listed in Table~\ref{table_TargetPropertiesTwo}. It shows the spectral
type, the number of finally used reference stars ($N_{\rm rs}$), the
$\chi^{2}$ value of the light curve and the one corresponding to $p=0.01$
($\chi^{2}_{0}$), variability flags, the probability of the null hypothesis
$p_{\rm tot}$, the variability amplitude, a possible period, the Amplitude
Ratio and the finally claimed variability flag. Here, the variability
amplitude is simply the root-mean-square (rms) of the relative light curve in
case of variable objects and an estimation of an upper limit above which
variability would have been detected for non-variable objects (using the
method of BJM01, Section~$5.1$). The variability flags in column $6$ state
whether or not a target is variable according to the $\chi^{2}$ test on the
individual observation nights (and in total in parentheses). In contrast to
this, the flag in the last column indicates whether we finally claim the
target to be variable, after consulting all means of investigation.
We claim a target to be generally variable if the probability of the
null hypothesis of the $\chi^{2}$ test is less than $1\,\%$. This
choice is quite arbitrary and using other limits would turn some
detections into non-detections or vice versa. Likewise, periodic
variability is claimed to be significant if the corresponding peak in
the periodogram has a FAP of less than $1\,\%$. In all cases where we
found variability or remarkable features in a target's light curves,
we also visually checked the reference stars' light curves,
variability measures and power spectra to rule out exterior
influences.

\subsection{Discussion of individual targets}\label{Results_Individual}

\subsubsection*{LHS370 (M5)}
\begin{figure}
\resizebox{\hsize}{!}{\includegraphics{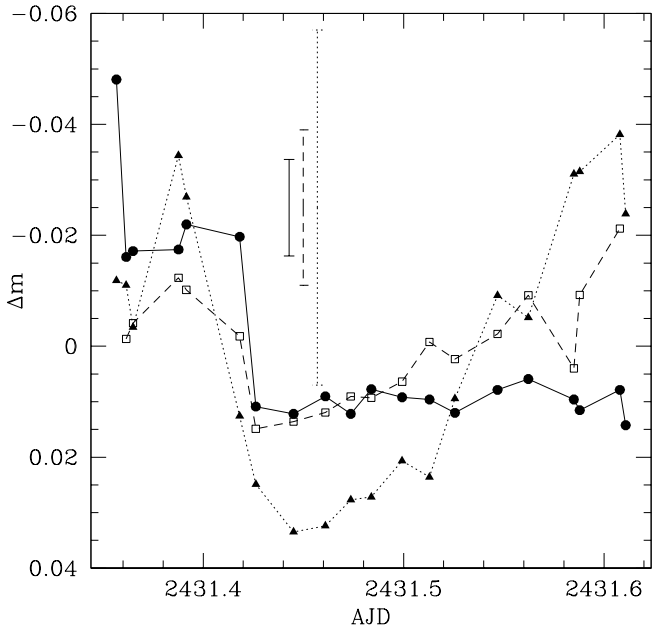}}
\caption{Relative light curves of LHS370 at an aperture radius of
nine pixels. The I, R and G band time series are shown with a solid, dashed
and dotted line, respectively. Typical error bars are plotted for each
channel.}
\label{figure_LHS370RelLC}
\end{figure}

\begin{figure}
\resizebox{\hsize}{!}{\includegraphics{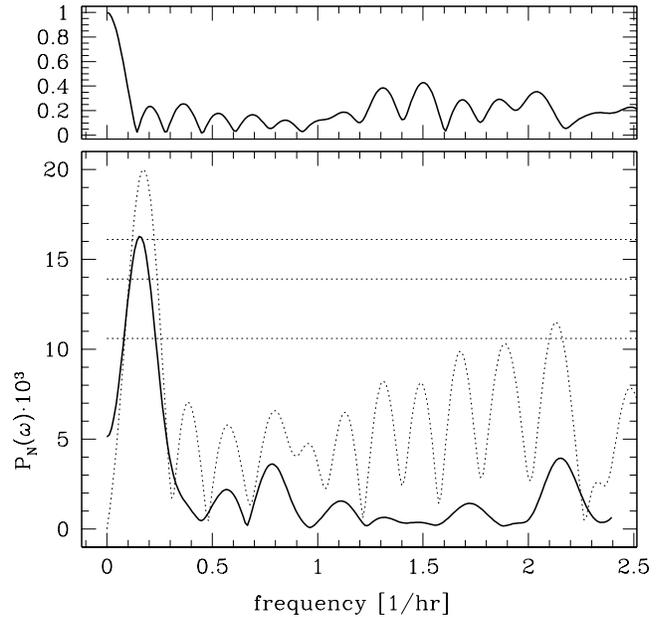}}
\caption{CLEAN periodogram of the G-band data of LHS370. The upper panel
  shows the spectral window function and the lower panel the dirty power
  spectrum as a dotted line as well as the cleaned power spectrum as a thick
  solid line. The $0.1\,\%$, $1\,\%$ and $10\,\%$ FAP power levels are plotted
  as dotted horizontal lines (from top to bottom).}
\label{figure_LHS370Clean_G}
\end{figure}

\begin{figure}
\resizebox{\hsize}{!}{\includegraphics{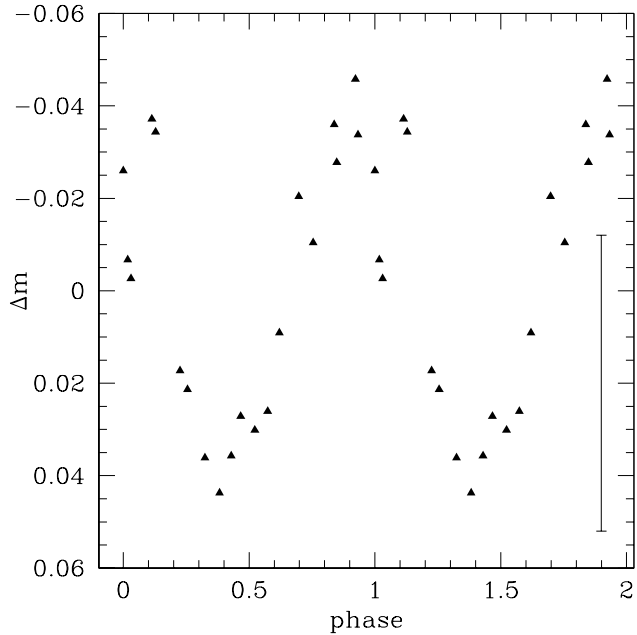}}
\caption{G band light curve of LHS370 phased to a period of
$6.5$ hours. The typical error is also shown. 
Note that two cycles of the phased light curve are plotted.
}
\label{figure_LHS370PhasedLC_G}
\end{figure}

Due to bad seeing of up to $3\arcsec$ on night $8$ in 2002, the star
aperture radius had to be increased for LHS370 from $6$ to $9$ pixels to
maximise the SNR. This object shows signs of variability in all three
channels. In I it is a strong detection ($p<10^{-6}$) according to the
$\chi^{2}$ test but a non-detection in R and G. Nevertheless, the light curves
(Fig.~\ref{figure_LHS370RelLC}) of R and G show
variations at similar amplitudes to that in I, but with larger errors. Although our data
only cover about one cycle, the shape of the light curves suggests a periodic
behaviour except for the I-band. This is confirmed since the power spectra do
contain significant (above the $0.1\,\%$ FAP power level) peaks in R and G at
periods of $5.9\pm 2.0\,{\rm hr}$ (theoretical uncertainty $2.9\,{\rm hr}$)
and $6.5\pm 2.0\,{\rm hr}$ ($\pm 3.5\,{\rm hr}$), respectively, but none
in~I. See Fig.~\ref{figure_LHS370Clean_G} for the CLEAN periodogram of the
G-band data and its light curve phased to the just mentioned period in
Fig.~\ref{figure_LHS370PhasedLC_G}.
Since the phased light curves in both R and G look reasonable, there
is good evidence for periodic variability. Studying the results of
the period uncertainty simulations (see above), we find that they are
consistent with a single periodicity. Since the Amplitude Ratio (AR) of the
I-band data is $2.4$ we can confidently infer that there is no
similar period in the I-band compared to those in R and
G. Because of the higher AR in R (compared to G), we claim the R-band
period at $5.9\pm 2.0\,{\rm hr}$ to be the object's rotation period. This
is consistent with $v\sin i$ measurements of various
M~dwarfs by \citet{Mohanty2003} and \citet{Delfosse1998}. To
investigate possible sources for this kind of variability, we compared
the observations with a grid of synthetic atmosphere spectra. For
details on this procedure see the section on 2M1707+64 below. From such an
analysis we could not match the observed variability amplitudes
simultaneously in the three channels with any reasonable atmospheric
model spectra featuring either magnetically induced star spots or dust
clouds. Alternative models for dust clouds or spots may be required.
We define the coverage factor $e$ as the fraction of the visible hemisphere
that is covered by surface features. By simply comparing the predicted
amplitudes with the observed, we can however place an upper limit of $e=0.05$
on the coverage factor of spots on the surface of LHS370. Otherwise we would
have been able to clearly detect their signature in the I-band. Higher
SNR data would yield a higher photometric precision and thus allow us
to better constrain the source of variability in LHS370.

\subsubsection*{2M1707+64 (M9)}
\begin{figure}
\resizebox{\hsize}{!}{\includegraphics{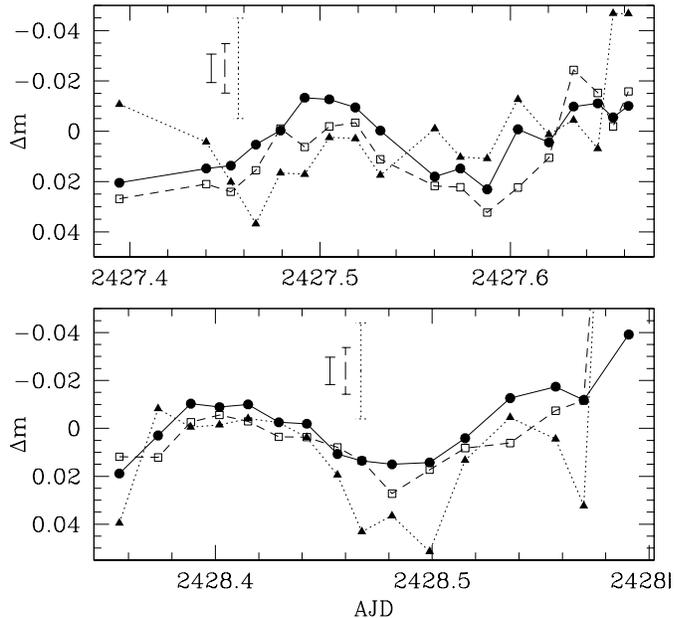}}
\caption{The relative light curves of 2M1707+64 in I (solid circles
  on a solid line), R (open squares on a dashed line) and G (solid triangles
  on a dotted line). Each observation night is shown in an individual panel
  with typical error bars for each channel.}
\label{figure_2M1707+64RelativeLightCurves}
\end{figure}

\begin{figure}
\resizebox{\hsize}{!}{\includegraphics{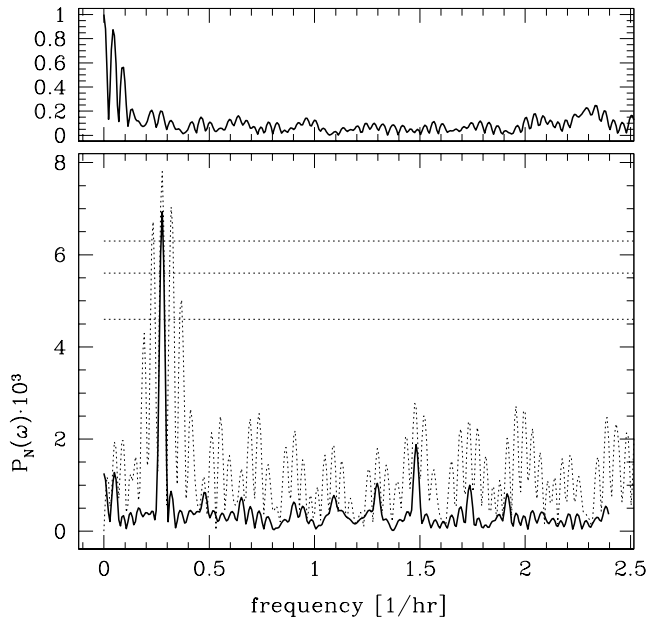}}
\caption{CLEAN periodogram of the I band light curve of 2M1707+64. The
  upper panels shows the spectral window function and the lower panel
  the dirty power spectrum as a dotted line as well as the cleaned
  power spectrum as a thick solid line. The $0.1\,\%$, $1\,\%$ and $10\,\%$
  FAP power levels are plotted as dotted horizontal lines (from top to
  bottom). Cf. Fig.~\ref{figure_LHS370Clean_G}. The above periodogram has
  higher resolution due to the longer time base.}
\label{figure_2M1707+64Clean_I}
\end{figure}

\begin{figure}
\resizebox{\hsize}{!}{\includegraphics{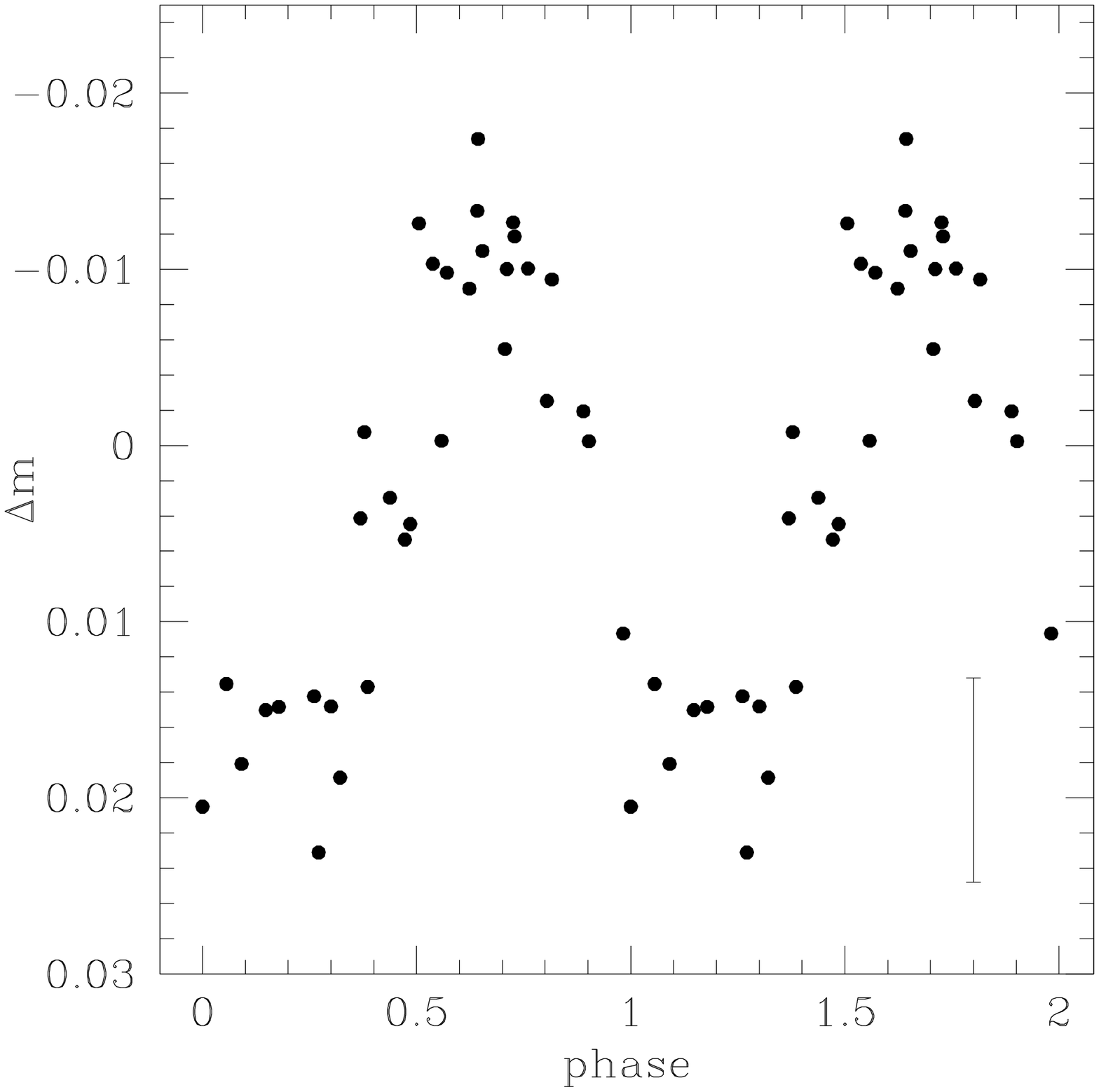}}
\caption{Relative light curve of 2M1707+64 in I phased to a period of
$3.65\,{\rm hr}$. The plot shows the combined data of both nights. The typical
error is also plotted. 
Note that two cycles of the phased light curve are plotted.}
\label{figure_2M1707+64PhasedLC_I}
\end{figure}

This target shows various kinds of variability. First, it is a strong
detection in terms of the $\chi^{2}$ test in I and R on both observation
nights ($p<10^{-6}$ and $p=0.001$, respectively) and only on night two in
G. Besides this general variability indication, periods with FAPs better than
$0.1\,\%$ at $3.65\,{\rm hr}$, $3.7\,{\rm hr}$ and $3.3\,{\rm hr}$ are
present in I, R and G respectively. The uncertainty in those periods is
estimated to be about $0.1\,{\rm hr}$ (theoretical estimate $0.6\,{\rm hr}$)
and thus all three values correspond to the same periodicity. As with LHS370,
this period is consistent with the $v\sin i$ values of late M type objects, so
we claim this to be the rotation
period. Figure~\ref{figure_2M1707+64RelativeLightCurves} shows that these
periods can be seen visually as nearly sinusoidal modulations in the light
curves. Figure~\ref{figure_2M1707+64Clean_I} shows the CLEAN power
spectrum in I and Fig.~\ref{figure_2M1707+64PhasedLC_I} the I band light curve
phased to the mentioned period.
A huge flare was detected at the end of the second night
which confirms that magnetic activity is significant in this late type
object. For more information on this and other flares found within
this data set, see \citet{Rockenfeller2006}.

Since the variations in the light curves of the different channels are
strongly correlated, it is obvious to assume that some surface feature
co-rotating with the target is the reason for the measured period. To
put stronger constraints on the source of variability in 2M1707+64, we
used the synthetic atmospheric models of \citet{Allard2001}. As shown
in \citet{Bailer-Jones2002}, it is straight forward to derive the
difference spectrum caused by co-rotating or forming / dissolving star
spots or dust clouds. By integrating the predicted spectral variations
over our photometric bands, we arrive at variability amplitudes for a
specific model (refer to the just mentioned paper for a description of
the two cases used, \emph{dusty} and \emph{cond}, as well as example
spectra plots). Since, for a specific model, the predicted amplitudes
depend linearly on the coverage factor $e$ of the surface features
(for values of up to $e=0.3$), the ratio of two amplitudes in different
wavebands will be independent of $e$. This makes these ratios suitable
for assessing whether or not an individual model fits our data. Based on
this, we built a grid of results for various trial atmospheric
models. The best-fitting model (dusty atmosphere with $T_{\rm
eff}=2300\,{\rm K}$, $\log g=5.0$ and a $100\,{\rm K}$ cooler spot)
along with the observational amplitudes are shown in
Table~\ref{table_Signatures_2M1707+64}, whereas
Table~\ref{table_Signatures_2M1707+64_II} shows some other results
from the computed model grid.
\begin{table*}
\caption{Comparison of observed and simulated variability
  amplitudes. The upper panel shows the observational standard (std),
  i.e.\ RMS, and peak-to-peak (ptp) amplitudes and the ratios formed
  hereof. Ratio errors are estimated from the errors in the relative
  magnitudes. Theoretical amplitudes and ratios are presented in the
  lower part. The four cases are cool spot or dusty cloud on a cond
  (clear) atmospheres, and cool spot or a hole (cond ``cloud'') on a
  dusty atmosphere. The underlying atmosphere has an effective
  temperature of $2300\,$K and surface gravity of $\log g=5.0$; the
  cool spot is $100\,$K cooler and all feature types (spot or cloud)
  have a projected surface coverage factor of $0.1$. These are the parameters
  of the best-fitting model. Values in parentheses indicate the
  amplitude ratios as explained in the last column.  Note that the
  amplitude ratios are independent of the coverage factor for small
  ($<0.3$) coverage factors.}
\label{table_Signatures_2M1707+64}
\centering
\begin{tabular}{lllllll}
\hline
\hline
\multicolumn{7}{c}{observational quantities}\\
band / colour  &  std ampl. [mag]  &  ptp ampl. [mag]  &  ratio  &
std ratio  &  ptp ratio  &  ratio error\\
\hline
G  &  0.022  &  0.055  &  $G:R$  &  1.57  &  \textbf{1.49}  &  $\pm 0.25$\\
R  &  0.014  &  0.037  &  $G:I$  &  1.83  &  \textbf{1.67}  &  $\pm 0.25$\\
I  &  0.012  &  0.033  &  $R:I$  &  1.17  &  \textbf{1.12}  &  $\pm 0.30$\\
\\
\hline
\hline
\multicolumn{7}{c}{theoretical quantities}\\
 &  \multicolumn{2}{c}{cond atmosphere}  &  \multicolumn{2}{c}{dusty
   atmosphere}  &  \multicolumn{2}{c}{legend}\\
band / colour  &  cool spot  &  dusty cloud  &  cool spot  &  cond cloud  &
\multicolumn{2}{c}{ampl. (ratio)}\\
\hline
G  &  0.039 (1.56)  &  0.064 (1.42)  &  0.072 (\textbf{1.48})  &  -0.135
(1.85) &  \multicolumn{2}{c}{G ($G:R$)}\\
R  &  0.025 (1.56)  &  0.045 (3.05)  &  0.049 (\textbf{1.70})  &  -0.073
(5.40) &  \multicolumn{2}{c}{R ($G:I$)}\\
I  &  0.025 (1.00)  &  0.021 (2.14)  &  0.042 (\textbf{1.15})  &  -0.025
(2.92) &  \multicolumn{2}{c}{I ($R:I$)}\\
\hline
\end{tabular}
\end{table*}
\begin{table*}
\caption{Supplementary data (to
Table~\ref{table_Signatures_2M1707+64}) that shows the amplitude
ratios as calculated with model parameters given in the first two
columns (the surface gravity is fixed at $\log g=5.0$ since it does
not have a significant influence on the results; feature sizes are
independent of the coverage factor up to $e\approx 0.3$). \emph{cond
spot} stands for a cool cond spot on a cond background atmosphere and
\emph{dusty cloud} similarly denotes a dusty cloud on a cond
atmosphere.}
\label{table_Signatures_2M1707+64_II}
\centering
\begin{tabular}{lllllllllll}
\hline
\hline
\multicolumn{11}{c}{grid of theoretical quantities -- amplitude ratios}\\
\hline
\multicolumn{2}{c}{model parameters}  &  \multicolumn{4}{c}{cond atmosphere}
& &  \multicolumn{4}{c}{dusty atmosphere}\\
\cline{3-6}  \cline{8-11}\\
$T_{\rm eff} [K]  $  &  $(\Delta T)_{\rm spot}  [K]  $  & & $G:R$  &  $G:I$  &
$R:I$  & & & $G:R$  &  $G:I$  &  $R:I$\\
\hline
2300  &  100  &  cond spot  &  1.56  &  1.56  &  1.00  & & dusty spot  &  1.48
&  1.70  &  1.15\\
      &       &  dusty cloud  &  1.42  &  3.05  &  2.14  & & cond cloud  &
1.85  &  5.40  &  2.92\\
2300  &  200  &  cond spot  &  1.62  &  1.38  &  0.85  & & dusty spot  &  1.24
&  1.33  &  1.07\\
2300  &  300  &  cond spot  &  1.60  &  1.24  &  0.77  & & dusty spot  &  1.14
&  1.16  &  1.02\\
\hline
2500  &  100  &  cond spot  &  1.22  &  1.92  &  1.57  & & dusty spot  &  1.27
&  1.96  &  1.54\\
      &       &  dusty cloud  &  1.44  &  6.82  &  4.73  & & cond cloud  &
1.59  &  8.70  &  5.46\\
2500  &  200  &  cond spot  &  1.22  &  1.70  &  1.39  & & dusty spot  &  1.21
&  1.66  &  1.38\\
2500  &  300  &  cond spot  &  1.22  &  1.49  &  1.22  & & dusty spot  &  1.15
&  1.38  &  1.20\\
\hline
2100  &  100  &  cond spot  &  2.25  &  1.16  &  0.51  & & dusty spot  &  1.36
&  1.23  &  0.91\\
      &       &  dusty cloud  &  1.21  &  1.60  &  1.33  & & cond cloud  &
2.57  &  5.50  &  2.14\\
\hline
\end{tabular}
\end{table*}
The theoretical predictions were made for cooler spots on either type
of background atmosphere (cond and dusty) as well as for clouds of the
opposite type and the same temperature as the background atmosphere,
i.e.\ a cond cloud on a dusty atmosphere (clear hole in a dusty sky)
and the other way around (dusty cloud on a clear sky). This setup excludes
fast convection as a source for the clouds since they would then be
hotter than the surrounding regions. Because of their small size
(compared to the noise in individual measurements), the determination
of the observational amplitudes is non-trivial. We decided to form
both the peak-to-peak amplitudes (ptp) after an exclusion of obvious
outliers and the root-mean-square of the light curve, here called
\emph{standard} amplitude (std). In the case of 2M1707+64, the
amplitude ratios formed with std and ptp amplitudes are similar
because of the nearly sinusoidal shape of the light curves. For a
perfectly sinusoidal signal the std amplitude is $\sqrt{\pi}/2$ times
the ptp amplitude.

Studying both tables for $T_{\rm eff}=2300\,{\rm K}$ and $T_{\rm
eff}=2500\,{\rm K}$, we can clearly exclude clouds as the source of
variability because the predicted signatures are too far from the measured
ones. The case of a dusty cloud on a $T_{\rm eff}=2100\,{\rm K}$ cond
atmosphere also is an acceptable fit to the data. But first, the effective
temperature is somewhat lower than the one derived by the spectral type
(\citet{Gorlova2003}, although the uncertainty in $T_{\rm eff}$ is about
$200\,{\rm K}$) and second, and more convincingly, cond atmospheres are
expected to occur not earlier than T-type objects. 

Besides the best-fitting one, models with spots on (particularly dusty)
atmospheres of $2200\leq T_{\rm eff}\leq 2500$ fit the observed amplitude
ratios within the estimated errors. Hence we conclude that co-rotating
magnetically induced spots, most likely on a dusty atmosphere of 
$2200-2500\,{\rm K}$, cause the observed variability in 2M1707+64. We infer
from the observed amplitudes that the coverage factor of these spots cannot be
larger than about $e=0.075$ for a dusty atmosphere (and about $0.15$ for a cond
one). Note that a coverage factor of e.g. $e=0.1$ does not necessarily refer
to a single spot covering ten percent of the visible hemisphere. One can
imagine other spot configurations leading to the same spectral signature as
that of a single spot. For example, a symmetrical distribution of many small
spots (with the same $e$) will not result in any detectable photometric
signal. Thus the value of $e=0.075$ found for 2M1707+64 could also mean that a
larger fraction of the star's hemisphere is covered by symmetrically
distributed spots and that there is a non-symmetric spot coverage of about
$7.5\,\%$.

\subsubsection*{CTI1709+27 (M5.5)}
\begin{figure}
\resizebox{\hsize}{!}{\includegraphics{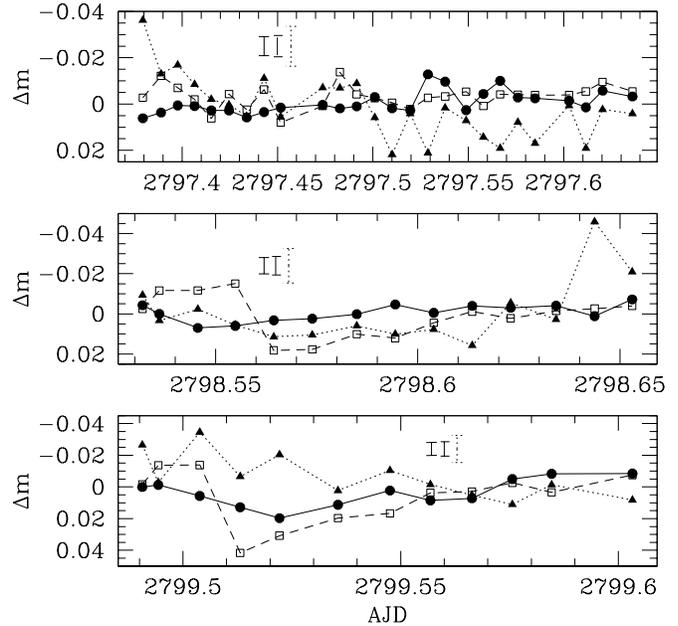}}
\caption{Relative light curves of CTI1709+27 in the I-band (solid line),
R-band (dashed line) and G-band (dotted line). Each observation night is shown
in an individual panel with typical error bars for each channel.}
\label{figure_CTI1709+27RelativeLightCurves}
\end{figure}

We present the relative light curves of this target in
Fig.~\ref{figure_CTI1709+27RelativeLightCurves} as an example for 
a typical object showing general variability above a probability of $99\,\%$
(here even at $p<10^{-6}$, see Table~\ref{table_TargetPropertiesTwo})
according to the $\chi^{2}$ test. No remarkable features or periodicities are
present.

The choice of the probability at which to claim a target to be
variable in terms of the $\chi^{2}$ test is quite arbitrary. The use
of $p=0.05$ instead of $p=0.01$ would have led to $7$ more targets to
be claimed variable (at least in one channel each)! Thus one should
first use a fairly low value for $p$ and second, it would be ideal to
re-monitor those targets after some time to test for persistent
variability. Of course, surface features could appear only temporarily
and thus lead to a non-detection during the re-monitoring, even though
they actually were present on the first observation run. Finally, this
problem can be overcome by higher SNR data that permit even lower
amplitude detections.

\subsection*{Further variable objects}
Hints of periodic variability were also found in 2M1344+77 in G
($12.5\pm 3.0\,{\rm hr}$), in LHS2930 in I ($13.2\pm 1.9\,{\rm hr}$)
and in 2M1714+30 in R ($6.9\pm 0.15\,{\rm hr}$) by the evaluation of
the CLEAN power spectra, although the the visual appearance of the
phased light curves do not support the presence of such periods. On
the other hand, simulations of periodic modulations show that we should
not always expect this, especially with multiple spots: see section~6.2 of
BJM01. Hence Table~\ref{table_TargetPropertiesTwo} lists these periods with a
question mark as these periods are somewhat tentative. Of these three
targets only 2M1714+30 (in R) is a detection according to the
$\chi^{2}$ test.

General variability according to the $\chi^{2}$ test was detected in
other targets as well, but they do not show any remarkable
features. These targets are: CTI1539+28 (I), 2M1546+37 (R), CTI1629+28
(R and G), LHS3307 (R), LHS3339 (I and G), LHS3376 (I). Variability in
the G-band data of CTI1629+28 was rejected because the two targets
which were also observed in the same nights show similar trends in
their corresponding G-band light curves. Because of bad data quality
in the case of LHS3376, the variations in I also cannot be considered
to be significant.

Significant correlations between different channels could only be
found for 2M1707+64 and LHS370, where they are obvious. Correlations
are another indicator of the reliability of a variability. Of course,
the underlying physical process could produce polychromatic variations
such that the signal is too small to be detected in one or more
channels.

In total, we find non-periodic variability in $5$ out of $19$ M~dwarfs
(or $4$, if we consider the period of 2M1714+30 to be real).  Periodic
variability was convincingly found in $2$ targets (or up to $5$, if we
choose to be less conservative). Although these results are surely
affected by small number statistics, they suggest that about
$2/7~\simeq~30\,\%$ of variable field M~dwarfs show periodic
variability.

The fact that we did not detect any variability for $12$ out of $19$
targets means that either no detectable surface features are present
on these objects or that their rotation periods are too long to be
detected. The latter may be particularly relevant for the four
earliest type objects in our sample with spectral types M2--M4, since
rotation periods of up to a few days are possible. The large fraction
of non-periodic variable targets ($5$ out of $7$ in total) could be
due to rapidly evolving surface features which would veil the rotation
period (the ``masking hypothesis'' of BJM01).

It is interesting to note that in the case of LHS370, no periodic
variability was found in the I-band even though we assign a rotation
period to this target on account of the R- and G-band data. Hence we
would have counted LHS370 to be generally variable but non-periodic if
it were only observed in the I-band. Since almost all recent
monitoring programs on UCDs have been conducted only in the I-band, it
is possible that the large number of non-periodic variable L dwarfs
found would show periods at other wavelengths. Multichannel
observations of L dwarfs are required to investigate this.

\subsection*{Non-variable objects}
All other targets show no signs of variability or exceptional
behaviour and are hence not mentioned individually. At the present
sensitivity limit, we consider them non-variable. See
Table~\ref{table_TargetPropertiesTwo} for upper limits on variability
amplitudes and for ARs. The latter allows us to judge whether we could
have missed existing periods because of this, see
Section~\ref{DetectionFraction}.

\subsection{Follow-up observations of 2M1707+64}
\begin{figure}
\resizebox{\hsize}{!}{\includegraphics{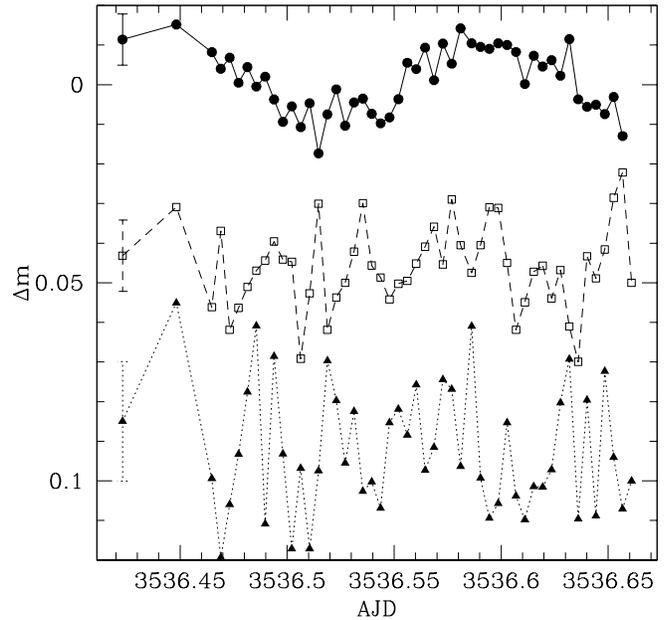}}
\caption{The relative light curves of the first night of
  follow-up observations of 2M1707+64 with BUSCA in June 2005. I-band data is
  represented by solid lines, R-band data by dashed lines and G-band data by
  dotted lines. Error bars are shown only for one data point per channel.}
\label{figure_2M1707+64RelLC2005_N01}
\end{figure}

\begin{figure}
\resizebox{\hsize}{!}{\includegraphics{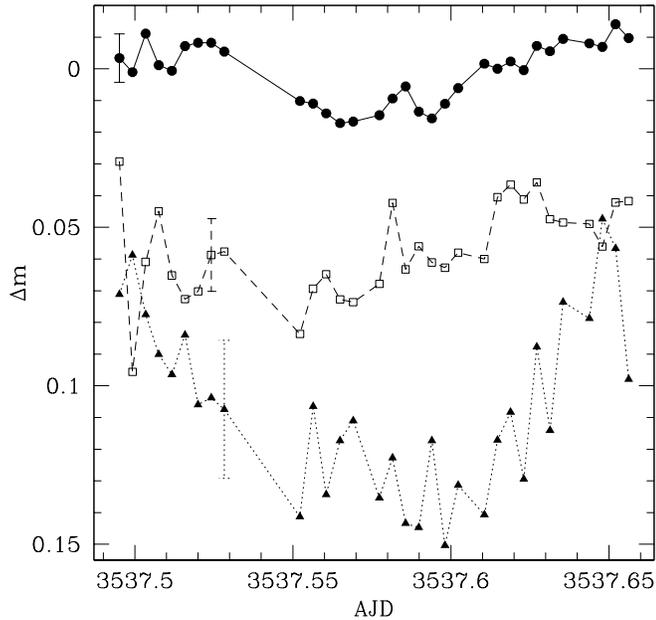}}
\caption{The relative light curves of the second night of
  follow-up observations on 2M1707+64 with BUSCA in June 2005. I-band data is
  represented by solid lines, R-band data by dashed lines and G-band data by
  dotted lines. Error bars are shown only for one data point per channel.}
\label{figure_2M1707+64RelLC2005_N02}
\end{figure}

Follow-up observations on 2M1707+64, also using the BUSCA instrument, were
performed as Directors Discretionary Time observations on the 14th and 15th of
June 2005. The data reduction and analysis were done in the same way as for
the 2002 and 2003 data. The relative light curves of I, R and G are shown in
Fig.~\ref{figure_2M1707+64RelLC2005_N01} and
Fig.~\ref{figure_2M1707+64RelLC2005_N02}.
All three channels are detections according to the $\chi^{2}$
test. Although the R- and G-band data seems to consist of large
random-like variations, the I-band light curve strongly resembles that
from the 2002 data. Moreover, the I-band power spectrum contains a
peak above the $0.1\,\%$ FAP power level at $3.61\pm 0.2\,{\rm hr}$
which confirms the one we detected in the earlier data. This period
can also be found on night two of the G-data ($P=3.80\pm 0.2\,{\rm
hr}$) at the same significance and on night one with $P=4.0\pm
0.27\,{\rm hr}$ at a FAP of $1\,\%$.  The constancy of the period
suggests that this is indeed a rotation period. We perhaps cannot say
{\it the} rotation period as the object may show differential
rotation, although if so, the spots would have to be at similar
latitudes (and/or the differential rotation would have to be
small). Due to the lack of a detected period in R, we could not repeat
the model spectra procedure to investigate the source of variability.

Further independent observations on 2M1707+64 were carried out at the
Maidanak Observatory, Uzbekistan, in the I-band in June 2005, as
reported in \citet{Rockenfeller2006}. These observations were made
primarily to further investigate the flare activity of this target,
but they also strongly support the rotation period, detected in the
Maidanak data at $3.63\pm 0.08\,{\rm hr}$.

\subsection{Variability dependences}\label{VariabilityDependences}
\begin{figure}
\resizebox{\hsize}{!}{\includegraphics{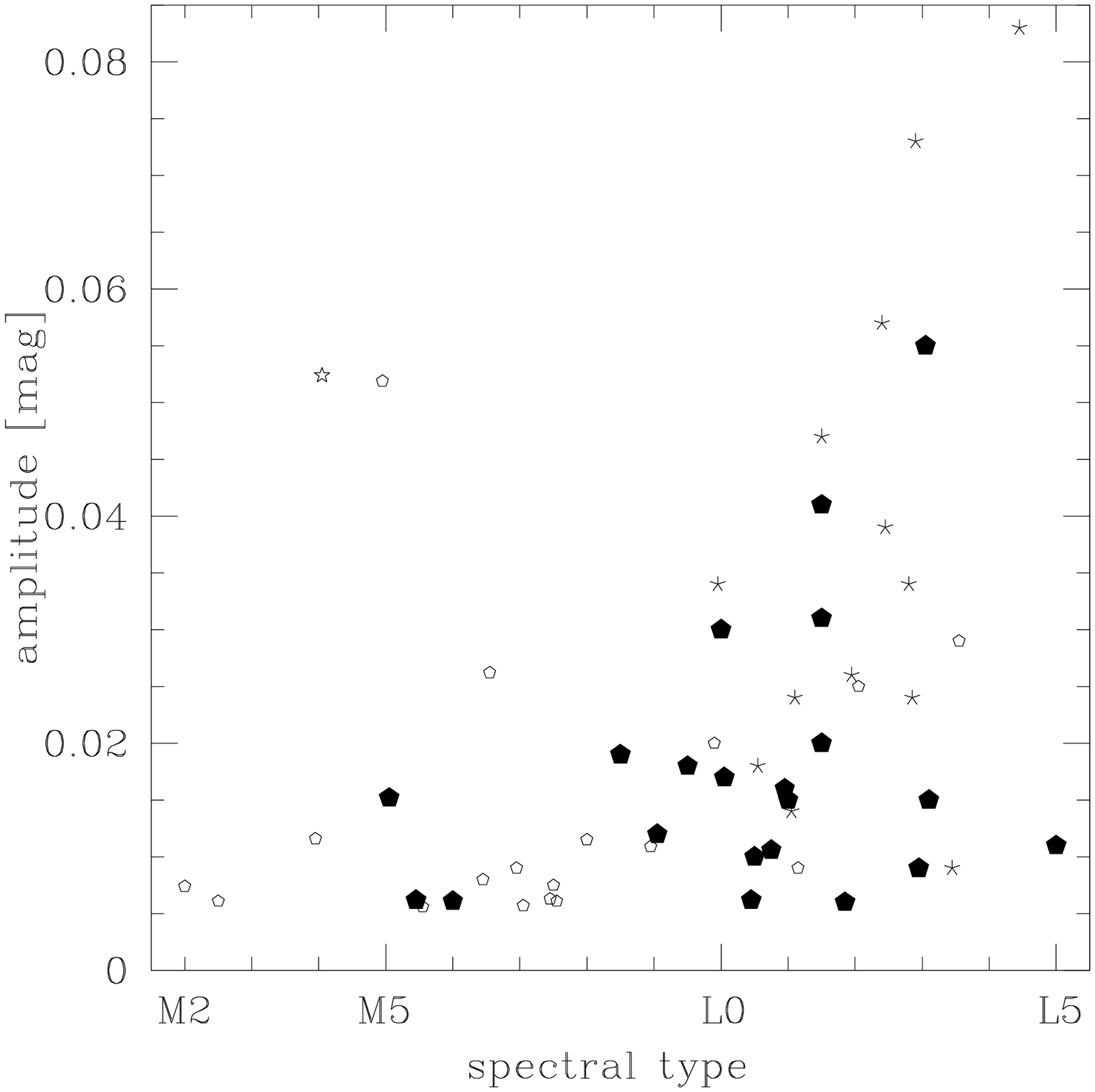}}
\caption{Variability amplitudes (variable objects, solid markers) and
  upper detection limits (non-variable objects, open markers) compiled from
  all relevant works on field dwarfs in the I-band of the recent years,
  including: this paper, \citet{Bailer-Jones2001}, \citet{Clarke2002},
  \citet{Clarke2003}, \citet{Gelino2002}, \citet{Koen2005},
  \citet{Martin2001}. The spectral types are slightly offset from the real
  ones in case there is more than one object of the same type. Amplitudes and
  upper limits are always rms values. Since \citet{Clarke2002} give peak-to-peak
  amplitudes for their two variable targets, we converted those to rms values
  assuming a sinusoidal signal (that assumption is not entirely justified, but
  differences in amplitudes are less than fifteen percent). Skeletal markers
  indicate rms amplitudes of non-variable objects where no upper limits are
  available. Starred symbols stand for rms-amplitudes where the $\chi^{2}$
  test gives a detection but variability is finally rejected.}
\label{figure_AmplitudesTotal}
\end{figure}

One of the major goals of this survey was to study whether variability is more
common in L dwarfs than in M dwarfs. To judge this, we compiled a list of
recently observed (1999--2005) M and L field dwarfs from various publications
(for sources, see the caption of Fig.~\ref{figure_AmplitudesTotal}). 
\begin{figure}
\resizebox{\hsize}{!}{\includegraphics{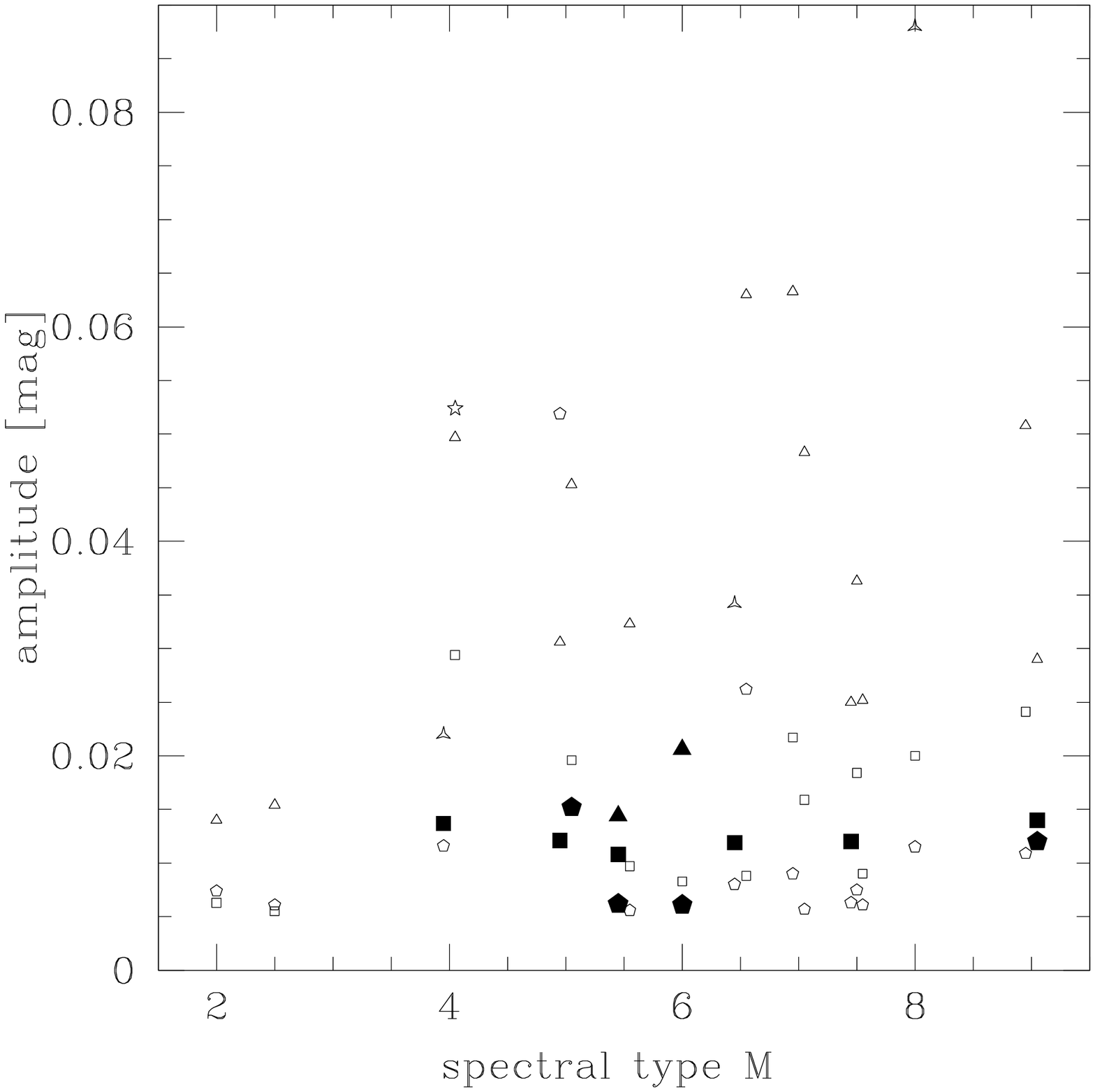}}
\caption{\small Variability amplitudes (variable objects, solid markers) and
  upper detection limits (non-variable objects, open markers) of our data set
  versus M-spectral type. Channels G, R and I are represented by triangles,
  squares and pentagons, respectively. Starred symbols stand for
  rms amplitudes where the $\chi^{2}$ test gives a detection but variability
  is finally rejected.}
\label{figure_AmplitudesPaper}
\end{figure}
A direct comparison of data published by different groups is ambiguous
since different detection limits were achieved or different
significance levels were used. We therefore only use those data which
we judge to be comparable to ours.  Fig.~\ref{figure_AmplitudesTotal}
shows the I band variability amplitudes (or upper limits for
non-variable objects, if available) versus the spectral type for these
samples.  Fig.~\ref{figure_AmplitudesPaper} shows variability
amplitudes and upper limits of the I-, R- and G-band data for all $19$
objects of our data set. No definite trend of the amplitudes with spectral
type is apparent in either figure, although amplitudes of more than
$0.02\,{\rm mag}$ have been found only at spectral types L0 to L3.

As most surveys have only been carried out in the I band, we use just
our I band variability detections when comparing our results with
other surveys.  We then form three spectral type bins -- early M
(M0--M4, $4$ objects), late M (M5--M9, 17 objects) and L (31 objects)
-- and calculate the fraction of variable objects in each. We thus
arrive at fractions of $0.25\pm 0.25$, $0.29\pm 0.13$ and $0.48\pm
0.12$ for these bins respectively (and $0.29\pm 0.12$ for all M dwarfs
together). Errors are derived assuming counting statistics. The
difference between M and L type is about $1.5\,\sigma$ which
corresponds to a confidence level of about $86\,\%$. Of particular
interest is to compare only the L~dwarfs from BJM01 with our M~dwarfs because
the analysis methods and sensitivity are very similar. We now find fractions
of $0.21\pm 0.11$ for the 19 M~dwarfs and $0.70\pm 0.26$ for the 10 L~dwarfs,
which is a $2\,\sigma$ difference. Both results suggest an increased frequency
of occurrence of variability among L dwarfs. This is supported further by the
fact that the present survey is more likely to detect variability in the M
dwarfs on account of the multichannel observations.

On account of the incompleteness of further information on the targets
as e.g.\ H$_{\alpha}$ or $v\sin i$, we cannot properly test whether
these quantities are significantly different between variable and
non-variable objects.


%% file: Conclusions.tex
\section{Conclusions}\label{Conclusions}
We have presented multiband data of 19 M dwarfs of which seven show
evidence for variability at a $99\,\%$ confidence level in at least
one of four channels. We performed relative photometry along with a
careful error estimation in order to achieve the high photometric
precision needed to detect low amplitude variability.  Amplitudes
(root-mean-square of the light curve) measured from $0.0055$ to
$0.014$ mag in I and R and from $0.014$ to $0.034$ mag in G.  For
non-variable objects, upper limits to variability are estimated to lie
between $0.006$ and $0.05$ mag.

Using the CLEAN algorithm to form the periodograms, convincing periodic
variability was found in LHS370 and 2M1707+64 at periods of $5.9\pm 2.0$ and
$3.65\pm 0.1$ hours, respectively. We further claim these to be their rotation
periods. Besides these two targets, three other objects show periodic
variations in one channel: 2M1344+77 in G at $12.5\pm 2.0\,$hr; LHS2930 in I
at $13.2\pm 1.9\,$hr; 2M1714+30 in R at $6.9\pm 0.15\,$hr. But here the
evidence for periodicity is less convincing.

Various simulations were performed to determine both the period
uncertainty and the probability of detecting an existing periodic
signal for arbitrary time samplings. In this way we could confirm the
high sensitivity of our data to the expected rotation periods of M
dwarfs up to about 12 hours (although early M~dwarfs may have longer
periods). With the help of synthetic spectra of \citet{Allard2001} the two
favourite scenarios for variability in ultra cool dwarfs -- magnetically
induced spots and dust clouds -- were studied in the cases of 2M1707+64 and
LHS370. It turned out that the observed variability in the former object can
be reproduced well by assuming an atmosphere of $2300\,$K, $\log g=5.0$ and a
$100\,$K cooler spot covering $7.5\,\%$ of the visible star disk. Clear
clouds on a dusty atmosphere or dusty clouds on a clear atmosphere can be
ruled out. In contrast to this, the origin of variability in LHS370
remains unclear. The lack of detectable periods for most of our
targets can be explained by absent or rapidly evolving surface
features.

Combining the results of this and other works published in the
recent years leads to evidence for an increased variability frequency
in field L~dwarfs when compared to field M-dwarfs. The fractions of
variable objects are: early M~type (M0--M4) $0.25\pm 0.25$, late M~type
(M5--M9.5) $0.29\pm 0.13$ and L~type $0.48\pm 0.12$. The number of
objects in each bin is 4, 17 and 31, respectively. The difference
between M and L type is about $1.5\sigma$ which corresponds to an
$86\,\%$ confidence level. This is statistically not large but still
hints at a more frequent variability in later spectral type. If we
consider the homogeneous samples of this paper and BJM01, we find
fractions of $0.21\pm 0.11$ (19 M~dwarfs) and $0.70\pm 0.26$ (10 L~dwarfs)
which corresponds to a $2\sigma$ difference.

Higher SNR data would allow us to put stronger constraints on
variability sources, particularly if optical and infrared observations
were combined.  An increased number of monitored M, late L and even T
dwarfs could clarify the change of variability frequency with spectral
type, and thus permit stronger conclusions. And while difficult to
achieve because of the available telescope time, a denser and longer
time sampling would make it possible to recover longer periods which
are expected to be common among early M dwarfs.


%% file: Paper.bbl
\begin{thebibliography}{32}
\expandafter\ifx\csname natexlab\endcsname\relax\def\natexlab#1{#1}\fi

\bibitem[{Allard {et~al.}(2001)Allard, Hauschildt, Alexander, Tamanai, \&
  Schweitzer}]{Allard2001}
Allard, F., Hauschildt, P.~H., Alexander, D.~R., Tamanai, A., \& Schweitzer, A.
  2001, \apj, 556, 357

\bibitem[{Bailer-Jones(2002)}]{Bailer-Jones2002}
Bailer-Jones, C.~A.~L. 2002, \aap, 389, 963

\bibitem[{Bailer-Jones(2004)}]{Bailer-Jones2004}
Bailer-Jones, C.~A.~L. 2004, \aap, 419, 703

\bibitem[{Bailer-Jones(2005)}]{Bailer-Jones2005}
Bailer-Jones, C.~A.~L. 2005, Proceedings 13th Cool Stars Workshop, F. Favata,
  G.~A.~J. Hussain, B. Battrick (eds), ESA, SP-560, 429

\bibitem[{Bailer-Jones \& Lamm(2003)}]{Bailer-Jones2003}
Bailer-Jones, C.~A.~L. \& Lamm, M. 2003, \mnras, 339, 477

\bibitem[{Bailer-Jones \& Mundt(1999)}]{Bailer-Jones1999}
Bailer-Jones, C.~A.~L. \& Mundt, R. 1999, \aap, 348, 800

\bibitem[{Bailer-Jones \& Mundt(2001)}]{Bailer-Jones2001}
Bailer-Jones, C.~A.~L. \& Mundt, R. 2001, \aap, 367, 218

\bibitem[{Bessell(1991)}]{Bessell1991}
Bessell, M.~S. 1991, \aj, 101, 662

\bibitem[{Clarke {et~al.}(2002)Clarke, Oppenheimer, \& Tinney}]{Clarke2002}
Clarke, F.~J., Oppenheimer, B.~R., \& Tinney, C.~G. 2002, \mnras, 335, 1158

\bibitem[{Clarke {et~al.}(2003)Clarke, Tinney, \& Hodgkin}]{Clarke2003}
Clarke, F.~J., Tinney, C.~G., \& Hodgkin, S.~T. 2003, \mnras, 341, 239

\bibitem[{Cruz {et~al.}(2003)Cruz, Reid, Liebert, Kirkpatrick, \&
  Lowrance}]{Cruz2003}
Cruz, K.~L., Reid, I.~N., Liebert, J., Kirkpatrick, J.~D., \& Lowrance, P.~J.
  2003, \aj, 126, 2421

\bibitem[{Dahn {et~al.}(2002)Dahn, Harris, Vrba, Guetter, Canzian, Henden,
  Levine, Luginbuhl, Monet, Monet, \& Pier}]{Dahn2002}
Dahn, C.~C., Harris, H.~C., Vrba, F.~J., {et~al.} 2002, \aj, 124, 1170

\bibitem[{Delfosse {et~al.}(1998)Delfosse, Forveille, Perrier, \&
  Mayor}]{Delfosse1998}
Delfosse, X., Forveille, T., Perrier, C., \& Mayor, M. 1998, \aap, 331, 581

\bibitem[{Gelino {et~al.}(2002)Gelino, Marley, Holtzman, Ackerman, \&
  Lodders}]{Gelino2002}
Gelino, C.~R., Marley, M.~S., Holtzman, J.~A., Ackerman, A.~S., \& Lodders, K.
  2002, \apj, 577, 433

\bibitem[{Gizis {et~al.}(2000)Gizis, Monet, Reid, Kirkpatrick, Liebert, \&
  Williams}]{Gizis2000}
Gizis, J.~E., Monet, D.~G., Reid, I.~N., {et~al.} 2000, \aj, 120, 1085

\bibitem[{Gorlova {et~al.}(2003)Gorlova, Meyer, Rieke, \&
  Liebert}]{Gorlova2003}
Gorlova, N.~I., Meyer, M.~R., Rieke, G.~H., \& Liebert, J. 2003, \apj, 593,
  1074

\bibitem[{Henry {et~al.}(2004)Henry, Subasavage, Brown, Beaulieu, \&
  Jao}]{Henry2004}
Henry, T.~J., Subasavage, J.~P., Brown, M.~A., Beaulieu, T.~D., \& Jao, W.-C.
  2004, \aj, 128, 2460

\bibitem[{Kirkpatrick {et~al.}(1994)Kirkpatrick, McGraw, Hess, Liebert, \&
  McCarthy}]{Kirkpatrick1994}
Kirkpatrick, J.~D., McGraw, J.~T., Hess, T.~R., Liebert, J., \& McCarthy,
  D.~W.~J. 1994, \apjs, 94, 749

\bibitem[{Koen(2005)}]{Koen2005}
Koen, C. 2005, \mnras, 357, 1151

\bibitem[{Lamm {et~al.}(2004)Lamm, Bailer-Jones, Mundt, Herbst, \&
  Scholz}]{Lamm2004}
Lamm, M.~H., Bailer-Jones, C.~A.~L., Mundt, R., Herbst, W., \& Scholz, A. 2004,
  \aap, 417, 557

\bibitem[{Leggett {et~al.}(2000)Leggett, Allard, Dahn, Hauschildt, Kerr, \&
  Rayner}]{Leggett2000}
Leggett, S.~K., Allard, F., Dahn, C.~C., {et~al.} 2000, \apj, 535, 965

\bibitem[{Mart\'in {et~al.}(1999)Mart\'in, Delfosse, Basri, Goldman, Forveille,
  \& Zapatero~Osorio}]{Martin1999}
Mart\'in, E.~L., Delfosse, X., Basri, G., {et~al.} 1999, \aj, 118, 2466

\bibitem[{Mart\'in {et~al.}(2001)Mart\'in, Zapatero~Osorio, \&
  Lehto}]{Martin2001}
Mart\'in, E.~L., Zapatero~Osorio, M.~R., \& Lehto, H.~J. 2001, \apj, 557, 822

\bibitem[{Mohanty \& Basri(2003)}]{Mohanty2003}
Mohanty, S. \& Basri, G. 2003, \apj, 583, 451

\bibitem[{Mohanty {et~al.}(2002)Mohanty, Basri, Shu, Allard, \&
  Chabrier}]{Mohanty2002}
Mohanty, S., Basri, G., Shu, F., Allard, F., \& Chabrier, G. 2002, \apj, 571,
  469

\bibitem[{Reid {et~al.}(2002)Reid, Kirkpatrick, Liebert, Gizis, Dahn, \&
  Monet}]{Reid2002}
Reid, I.~N., Kirkpatrick, J.~D., Liebert, J., {et~al.} 2002, \aj, 124, 519

\bibitem[{Roberts(1987)}]{Roberts1987}
Roberts, D.~H. 1987, \aj, 93, 968

\bibitem[{Rockenfeller(2005)}]{Rockenfeller2005a}
Rockenfeller, B. 2005, Diploma thesis, University of Heidelberg, Germany;
  http://www.mpia.de/homes/calj/MV\_survey.html

\bibitem[{Rockenfeller {et~al.}(2006)Rockenfeller, Bailer-Jones, Mundt, \&
  Ibrahimov}]{Rockenfeller2006}
Rockenfeller, B., Bailer-Jones, C.~A.~L., Mundt, R., \& Ibrahimov, M. 2006,
  submitted to MNRAS

\bibitem[{Scholz \& Eisl\"offel(2004a)}]{Scholz2004a}
Scholz, A. \& Eisl\"offel, J. 2004a, \aap, 421, 259

\bibitem[{Scholz \& Eisl\"offel(2004b)}]{Scholz2004b}
Scholz, A. \& Eisl\"offel, J. 2004b, \aap, 419, 249

\bibitem[{West {et~al.}(2004)West, Hawley, Walkowicz, Covey, Silvestri,
  Raymond, Harris, McGehee, Ivezi\'c, \& Brinkmann}]{West2004}
West, A.~A., Hawley, S.~L., Walkowicz, L.~M., {et~al.} 2004, \aj, 128, 426

\end{thebibliography}
